\documentclass[aps,prd,onecolumn,floatfix,nofootinbib,showpacs]{revtex4-1}

\pdfoutput=1
\usepackage{graphicx,color,float}
\usepackage{hyperref}
\usepackage{tabularx}
\usepackage{multirow}
\usepackage{verbatim}
\usepackage{longtable}
 \usepackage{amsmath}
\usepackage{amsfonts}
\usepackage{amssymb}
\usepackage{rotating}
\usepackage[dvipsnames]{xcolor}

\usepackage{setspace}

\usepackage{ulem}

\newcommand{\imag}{\Im {\rm m}}
\newcommand{\real}{\Re {\rm e}}
\newcommand{\lsim}{\raisebox{-0.13cm}{~\shortstack{$<$ \\[-0.07cm] $\sim$}}~}
\newcommand{\gsim}{\raisebox{-0.13cm}{~\shortstack{$>$ \\[-0.07cm] $\sim$}}~}
\def\slash#1{#1\!\!\!/}

\begin{document}

{\small
\begin{flushright}
IUEP-HEP-22-02
\end{flushright} }

\title{
On the mass spectrum of heavy Higgs bosons in 
two-Higgs-doublet model\\
in the light of the CDF $W$-mass anomaly 
}

\def\slash#1{#1\!\!/}

\renewcommand{\thefootnote}{\arabic{footnote}}

\author{
Dong-Won Jung,$^{1,2}$\footnote{dongwon.jung@yonsei.ac.kr}~
Yongtae Heo,$^{1,3}$\footnote{yongtae1heo@gmail.com}~
and
Jae Sik Lee$^{1,3}$\footnote{jslee@jnu.ac.kr}
}

\affiliation{
$^1$ Department of Physics, Chonnam National University, 
Gwangju 61186, Korea\\
$^2$ Department of Physics, 
Yonsei University, Seoul 03722, Korea\\
$^3$ IUEP, Chonnam National University, Gwangju 61186, Korea 
}
\date{June 28, 2023}

\begin{abstract}
\begin{spacing}{1.30}
We analyze the mass spectrum of the charged and neutral Higgs bosons
in the framework of two Higgs doublet model (2HDM)
in the light of the precision measurement of
the $W$ boson mass by the CDF collaboration.
We have considered the most general 2HDM potential
with explicit CP violation in the Higgs basis which contains
the three CP-mixed neutral mass eigenstates $H_1$, $H_2$, and $H_3$
with $M_{H_1}\leq M_{H_2}\leq M_{H_3}$.
The high-precision CDF measurement of the $W$ boson mass
is characterized by the large positive value of the $T$ parameter.
By identifying the lightest neutral Higgs boson $H_1$
as the SM-like one discovered at the LHC,
we find that 
it is necessary to have the mass splitting between the 
charged Higgs boson $H^\pm$ and the second heaviest neutral one $H_2$
to accommodate the sizable positive
deviation of the $T$ parameter from its SM value of $0$.
By combining the mass splitting between $H^\pm$ and $H_2$ with
the theoretical constraints 
from the perturbative unitarity 
and for the Higgs potential to be bounded from below, we
implement comprehensive analysis of
the mass spectrum of the heavy Higgs bosons 
taking account  of the effects of deviation from the alignment limit
and also the mass splitting between $H_3$ and $H_2$.
We further analyze the behavior of the heavy-Higgs mass spectrum 
according to the variation of the $T$ parameter.
Finally, 
concentrating on the heavy mass region 
$M\equiv (M_{H_2}+M_{H^\pm})/2\gsim 500$ GeV in which 
we find a mass  hierarchy of 
$M\gg M_{H_1}\sim |\Delta| \gsim \delta$
with $\Delta \equiv M_{H_2}-M_{H^\pm}$ and
$\delta \equiv  M_{H_3}-M_{H_2}$,
we discuss some benchmarking scenarios 
for the searches of heavy Higgs bosons at future colliders such as
the high luminosity option of the LHC and
a 100 TeV hadron collider.
\end{spacing}
\end{abstract}

\maketitle

\section{Introduction}
\label{sec:introduction}
Recently, the CDF collaboration has reported the result of
a new $W$ boson mass measurement with unprecedented precision
\cite{CDF:2022hxs}
\begin{equation}
80,433.5 \ \pm \ 9.4 \ {\rm MeV}\,,
\end{equation}
using $8.8$ fb$^{-1}$
data collected at the Fermilab Tevatron collider
with a center-of-mass energy of 1.96 TeV.
Comparing with the SM expectation of
$80,357 \ \pm \ 6 \ {\rm MeV}$~\cite{ParticleDataGroup:2020ssz},
the new measurement of the $W$ boson mass
suggests a difference with a significance of $7.0\sigma$.
Performing a global fit of electroweak data together with
the high-precision CDF measurement while fixing $U=0$, one may find
the large central values
of the oblique parameters $S$ and $T$
together with the standard deviations such as
\cite{Lu:2022bgw}
\begin{equation}
\label{eq:STCDF}
(\widehat S_0\,,\ \sigma_S)\ =\ (0.15\,,\ 0.08)\;,\qquad
(\widehat T_0\,,\ \sigma_T)\ =\ (0.27 \,,\ 0.06)\; ,
\end{equation}
and a strong correlation $\rho_{ST}=0.93$ between them.

Previously, by analyzing the large deviation of the $S$ and
$T$ parameters from their SM values of zero
in the framework of two Higgs doublet model (2HDM),
we reported the existence of the upper limit of about $1$ TeV on the masses of the heavy 
charged and neutral Higgs bosons~\cite{Heo:2022dey}.

To be specific, we take the CP-conserving (CPC) 2HDM Higgs potential 
and identify the lighter CP-even neutral Higgs boson 
as the SM-like one discovered at the LHC.
Observing that the mass difference between the charged Higgs boson $H^\pm$
and the CP-odd neutral one $A$ is
required to achieve the sizable positive central value of the $T$ parameter,
we derive the upper limit on the masses of heavy 
Higgs bosons by combining the mass difference with 
the theoretical constraints on the quartic couplings of the Higgs potential
from the perturbative unitarity and
for the Higgs potential to be bounded from below.
Quantitatively, we present our estimation that $M_{H^\pm} \lsim 1,000^{-100}_{+400} $ GeV
taking $\widehat{T}_0=0.27\pm 0.06$ in the alignment limit 
where the lighter CP-even neutral Higgs boson 
behaves exactly as the SM Higgs boson~\cite{Heo:2022dey}.
We note that similar results have been obtained by the authors of
Ref.~\cite{Lee:2022gyf}.
For other relevant studies of the CDF $W$-mass anomaly in the 
framework of 2HDM, we refer to 
Refs.~\cite{Fan:2022dck,Lu:2022bgw,Song:2022xts,Bahl:2022xzi,
Babu:2022pdn,Ahn:2022xax,Han:2022juu,
Arcadi:2022dmt,Ghorbani:2022vtv,Lee:2022gyf,Benbrik:2022dja,Abouabid:2022lpg,
Botella:2022rte,Kim:2022xuo,Kim:2022hvh,Appelquist:2022qgl,
Atkinson:2022qnl,Hessenberger:2022tcx,Kim:2022nmm,Arco:2022jrt,Kang:2022mdy,Kumar:2022rcf}.

In this work, 
compared to our previous study~\cite{Heo:2022dey},
we extend our analysis by considering
the Higgs potential with explicit CP violation.
In the presence of CP-violating (CPV) phases in the Higgs potential,
one cannot specify the neutral Higgs bosons
according to their CP parities as in the CPC case and, instead, 
we have the three CP-mixed
mass eigenstates of $H_1$, $H_2$, and $H_3$.
Therefore, the observation made in the CPC case based
on the mass difference between the charged Higgs boson 
and the CP-odd neutral one does fail and one needs the corresponding
condition in the CPV case.
By identifying the lightest neutral Higgs boson $H_1$
as the SM-like one, we find that it is necessary to have
the mass splitting between the charged Higgs boson $H^\pm$ and
the second heaviest neutral Higgs boson $H_2$ to accommodate
the large positive value of the $T$ parameter.
And we improve our previous study taking account  of
the effects of deviation from the alignment limit
and the mass splitting between the heavy neutral Higgs bosons.
Through the more comprehensive and improved analysis,
we have fully figured out all the physics origins 
relevant for the mass scale of heavy Higgs
bosons and suggest the two specific types of the mass 
spectrum of them.
We also address the situation in which
the current CDF $W$-mass anomaly
is ameliorated to make our analysis more convincing and
our results less restrictive.

This paper is organized as follows.
Section~\ref{sec:framework} is devoted to a brief review of the 2HDM Higgs potential
in the Higgs basis with explicit CP violation.
In Section~\ref{sec:constraints}, 
we elaborate on the constraints from the perturbative unitarity,
the Higgs potential bounded from below, and the electroweak precision observables.
In Section~\ref{sec:analysis},
we provide the comprehensive analysis 
why the mass splitting between the charged 
and the second heaviest neutral Higgs bosons is necessary and
why the masses of the heavy Higgs bosons should be bounded 
from above.
In Section~\ref{sec:discussion},
we show how the mass scale of the heavy Higgs bosons behaves
according to the variation of
the central values of the $S$ and $T$ parameters and
propose some benchmarking scenarios 
for the searches of heavy Higgs bosons at future colliders.
A brief summary and conclusions are made in Section~\ref{sec:conclusions}.

\section{Framework}
\label{sec:framework}
The general 2HDM scalar potential in the so-called Higgs basis
\cite{Donoghue:1978cj,Georgi:1978ri}
where only one doublet contains the non-vanishing vacuum expectation value
$v$ is given by~\cite{Lee:2021oaj}
\begin{eqnarray}
\label{eq:VHiggs}
V_{\cal H} &=&
Y_1 ({\cal H}_1^{\dagger} {\cal H}_1)
+Y_2 ({\cal H}_2^{\dagger} {\cal H}_2)
+Y_3 ({\cal H}_1^{\dagger} {\cal H}_2)
+Y_3^{*}({\cal H}_2^{\dagger} {\cal H}_1) \nonumber \\
&&+ Z_1 ({\cal H}_1^{\dagger} {\cal H}_1)^2 + Z_2
({\cal H}_2^{\dagger} {\cal H}_2)^2 + Z_3 ({\cal H}_1^{\dagger}
{\cal H}_1)({\cal H}_2^{\dagger} {\cal H}_2) + Z_4 ({\cal H}_1^{\dagger}
{\cal H}_2)({\cal H}_2^{\dagger} {\cal H}_1) \nonumber \\
&&+ Z_5 ({\cal H}_1^{\dagger} {\cal H}_2)^2 +
Z_5^{*} ({\cal H}_2^{\dagger} {\cal H}_1)^2 + Z_6
({\cal H}_1^{\dagger} {\cal H}_1) ({\cal H}_1^{\dagger} {\cal H}_2) + Z_6^{*}
({\cal H}_1^{\dagger} {\cal H}_1)({\cal H}_2^{\dagger} {\cal H}_1) \nonumber \\
&& + Z_7 ({\cal H}_2^{\dagger} {\cal H}_2) ({\cal H}_1^{\dagger} {\cal H}_2) +
Z_7^{*} ({\cal H}_2^{\dagger} {\cal H}_2) ({\cal H}_2^{\dagger} {\cal H}_1)\;,
\end{eqnarray}
which contains 3 dimensionful quadratic and 7 dimensionless quartic parameters of which 
four parameters of $Y_3$ and $Z_{5,6,7}$ are complex. 
The complex SU(2)$_L$ doublets of ${\cal H}_1$ and ${\cal H}_2$
can be parameterized as
\begin{eqnarray}
\label{eq:H12inHiggBasis}
{\cal H}_1=
\left(\begin{array}{c}
G^+ \\ \frac{1}{\sqrt{2}}\,(v+\varphi_1+iG^0)
\end{array}\right)\,; \ \ \
{\cal H}_2=
\left(\begin{array}{c}
H^+ \\ \frac{1}{\sqrt{2}}\,(\varphi_2+ia)
\end{array}\right)\,,
\end{eqnarray}
where $v = \left(\sqrt{2}G_F\right)^{-1/2} \simeq 246.22$ GeV and
$G^{\pm,0}$ and $H^\pm$ stand for the
Goldstone and charged Higgs bosons, respectively.
For the neutral Higgs bosons, 
the CP-odd state $a$ and the two CP-even states
$\varphi_1$ and $\varphi_2$ result in
three CP-mixed mass eigenstates $H_{1,2,3}$ through mixing 
and one of them should play the role of the SM Higgs boson.
The tadpole conditions relate the quadratic parameters $Y_1$ and $Y_3$
to  $Z_1$ and $Z_6$, respectively, as follows:
\begin{equation}
\label{eq:higgsbasistadpole}
Y_1 \ + \ Z_1 v^2\ = 0 \,; \ \ \
Y_3 \ + \ \frac{1}{2}Z_6 v^2\ = 0 \,.
\end{equation}
The 2HDM Higgs potential includes the mass
terms which can be cast into the form consisting of two parts
\begin{equation}
V_{{\cal H}\,, {\rm mass}}=
M_{H^\pm}^2 H^+ H^- \ + \ 
\frac{1}{2}
(\varphi_1 \ \varphi_2 \ \ a )\,{\cal M}^2_0\,
\left(\begin{array}{c}
\varphi_1 \\ \varphi_2  \\ a \end{array}\right)\,,
\end{equation}
in terms of the charged Higgs bosons $H^\pm$,
the neutral CP-even Higgs bosons $\varphi_{1,2}$, and
the neutral CP-odd Higgs boson $a$.
The charged Higgs boson mass is given by
\begin{equation}
\label{eq:mch}
M_{H^\pm}^2= Y_2 +\frac{1}{2} Z_3 v^2\,, 
\end{equation}
while the $3\times 3$ mass-squared matrix of the neutral Higgs
bosons ${\cal M}_0^2$ takes the form
\begin{equation}
\label{eq:m0sq}
{\cal M}^2_0 = \left(\begin{array}{ccc}0  & 0 & 0 \\ 0 & M_A^2 & 0\\ 0 & 0 & M_A^2
\end{array}\right)
\ + \
\left(\begin{array}{ccc}
2 Z_1 & \real(Z_6) & -\imag(Z_6)  \\
\real(Z_6) & 2 \real(Z_5 ) & -\imag(Z_5) \\
-\imag(Z_6) & -\imag(Z_5) & 0
\end{array}\right)\,v^2\,,
\end{equation}
where $M_A^2=M_{H^\pm}^2+ \left[Z_4/2 -\real(Z_5)\right]v^2$.
Note that the quartic couplings $Z_2$ and $Z_7$ have nothing to
do with the masses of Higgs bosons and the mixing of the neutral ones.
We further note that $\varphi_1$ does not mix with
$\varphi_2$ and $a$ in the $Z_6=0$ limit
and $\varphi_2$ does not mix with $a$ 
if $\imag{Z_5}=0$ imposed additionally.
Therefore, when $Z_6=\imag{Z_5}=0$, the three states $\varphi_1$,
$\varphi_2$, and $a$ themselves constitute the three mass eigenstates
and $\varphi_1$ plays the role of the SM Higgs boson.

Given the $3\times 3$ real and symmetric mass-squared ${\cal M}_0^2$,
one may describe the mixing among the neutral Higgs bosons by introducing the
$3\times 3$ orthogonal matrix $O$
\begin{eqnarray}
\label{eq:aOi}
(\varphi_1,\varphi_2,a)^T_\alpha=O_{\alpha i} (H_1,H_2,H_3)^T_i\,,
\end{eqnarray}
which satisfies the relation
\begin{equation}
\label{eq:OTM2O}
O^T {\cal M}_0^2 O={\rm diag}(M_{H_1}^2,M_{H_1}^2,M_{H_3}^2)
\end{equation}
with the increasing ordering of $M_{H_1}\leq M_{H_2}\leq M_{H_3}$.
When the three masses of the neutral Higgs bosons and the elements of
the orthogonal mixing matrix $O$ are given, the quartic couplings
$Z_1$, $Z_4$, $Z_5$, and $Z_6$ in the neutral-Higgs-boson 
mass-squared matrix ${\cal M}_0^2$ are given by
\begin{eqnarray}
\label{eq:Z1456}
Z_1 &=& \frac{1}{2 v^2} \left(
M_{H_1}^2 O_{\varphi_11}^2 + M_{H_2}^2 O_{\varphi_12}^2 +
M_{H_3}^2 O_{\varphi_13}^2 \right) \,, \nonumber \\[2mm]
Z_4 &=& \frac{1}{v^2}
\left[M_{H_1}^2 (O_{\varphi_21}^2 + O_{a1}^2) +
M_{H_2}^2 (O_{\varphi_22}^2 + O_{a2}^2)+
M_{H_3}^2 (O_{\varphi_23}^2 + O_{a3}^2) - 2 M_{H^\pm}^2
\right] \,, \nonumber \\[2mm]
Z_5 &=& \frac{1}{2 v^2}
\left[M_{H_1}^2 (O_{\varphi_21}^2 - O_{a1}^2) +
M_{H_2}^2 (O_{\varphi_22}^2 - O_{a2}^2)+
M_{H_3}^2 (O_{\varphi_23}^2 - O_{a3}^2) \right]  \nonumber \\
&& -\frac{i}{v^2} \left(
M_{H_1}^2 O_{\varphi_21}O_{a1} + M_{H_2}^2 O_{\varphi_22}O_{a2}
+ M_{H_3}^2 O_{\varphi_23}O_{a3} \right) \,, \nonumber \\[2mm]
Z_6 &=& \frac{1}{v^2} \left(
M_{H_1}^2 O_{\varphi_11}O_{\varphi_21} + M_{H_2}^2 O_{\varphi_12}O_{\varphi_22} +
M_{H_3}^2 O_{\varphi_13}O_{\varphi_23} \right) \nonumber \\
&& -\frac{i}{v^2} \left(
M_{H_1}^2 O_{\varphi_11}O_{a1} + M_{H_2}^2 O_{\varphi_12}O_{a2} +
M_{H_3}^2 O_{\varphi_13}O_{a3} \right) \,,
\end{eqnarray}
for given $v$ and $M_{H^\pm}$.
We observe that $\imag(Z_5)$ and $\imag(Z_6)$ contain the products
of $O_{\varphi_2 i}O_{a i}$ and $O_{\varphi_1 i}O_{a i}$ with $i=1,2,3$, 
respectively, each of which identically vanishes if the mass eigenstate
$H_{i}$ carries its own definite CP parity.

Once the mixing matrix $O$ is given,
the cubic interactions of the neutral and charged Higgs bosons
with the massive vector bosons $Z$ and $W^\pm$ are described by
the following interaction Lagrangians:
\begin{eqnarray}
{\cal L}_{HVV} & = & g\,M_W \, \left(W^+_\mu W^{- \mu}\ + \
\frac{1}{2c_W^2}\,Z_\mu Z^\mu\right) \, \sum_i \,g_{_{H_iVV}}\, H_i
\,,\nonumber\\[3mm]
{\cal L}_{HHZ} &=& \frac{g}{2c_W} \sum_{i>j} g_{_{H_iH_jZ}}\, Z^{\mu}
(H_i\, \!\stackrel {\leftrightarrow} {\partial}_\mu H_j) \,, \nonumber\\[3mm]
{\cal L}_{HH^\pm W^\mp} &=& -\frac{g}{2} \, \sum_i \, g_{_{H_iH^+
W^-}}\, W^{-\mu} (H_i\, i\!\stackrel{\leftrightarrow}{\partial}_\mu
H^+)\, +\, {\rm h.c.}\,,
\end{eqnarray}
where $X\stackrel{\leftrightarrow}{\partial}_\mu Y
=X\partial_\mu Y -(\partial_\mu X)Y$ and $i,j =1,2,3$.
Note that the normalized couplings
$g_{_{H_iVV}}$, $g_{_{H_iH_jZ}}$ and $g_{_{H_iH^+
W^-}}$ are given in terms of the neutral Higgs-boson $3\times 3$
mixing matrix $O$ by (note that det$(O)=\pm1$ for any orthogonal matrix $O$):
\begin{eqnarray}
\label{eq:2hdmhvvetc}
g_{_{H_iVV}} &=& O_{\varphi_1 i}
\, ,\nonumber \\
g_{_{H_iH_jZ}} &=& {\rm sign} [{\rm det}(O)] \, \, \epsilon_{ijk}\,
g_{_{H_kVV}}\, = {\rm sign} [{\rm det}(O)] \, \, \epsilon_{ijk}\, O_{\varphi_1 k}\,,
\nonumber \\
g_{_{H_iH^+ W^-}} &=& -O_{\varphi_2 i} + i O_{ai}  \, ,
\end{eqnarray}
leading to the following sum rules:
\begin{equation}
\label{eq:2hdmsumrule}
\sum_{i=1}^3\, g_{_{H_iVV}}^2\ =\ 1\,\quad{\rm and}\quad
g_{_{H_iVV}}^2+|g_{_{H_iH^+ W^-}}|^2\ =\ 1\,\quad {\rm for~ each}~
i=1,2,3\,.
\end{equation}

Introducing the three mixing angles $\gamma$, $\eta$, and $\omega$,
the orthogonal mixing matrix $O$ might be parameterized as follows:
\begin{eqnarray}
\label{eq:omix}
O = O_\gamma O_\eta O_\omega &\equiv &
\left( \begin{array}{ccc}
  c_\gamma  &   s_\gamma   &  0   \\
  -s_\gamma  &   c_\gamma  &  0   \\
  0         &      0       &  1   \\
  \end{array} \right)
\left( \begin{array}{ccc}
    c_\eta               &      0             &   s_\eta  \\
    0                    &      1             &   0   \\
   -s_\eta               &      0             &   c_\eta   \\
  \end{array} \right)
\left( \begin{array}{ccc}
    1               &      0       &   0   \\
    0               &   c_\omega   &   s_\omega   \\
    0               &  -s_\omega   &   c_\omega   \\
  \end{array} \right)
\nonumber \\[3mm] &=&
\left( \begin{array}{ccc}
    c_\gamma c_\eta  &  s_\gamma c_\omega - c_\gamma s_\eta s_\omega   &
    s_\gamma s_\omega + c_\gamma s_\eta c_\omega     \\
    -s_\gamma c_\eta   &  c_\gamma c_\omega + s_\gamma s_\eta s_\omega   &
    c_\gamma s_\omega - s_\gamma s_\eta c_\omega     \\
    -s_\eta  &  -c_\eta s_\omega   &    c_\eta c_\omega  \\
  \end{array} \right)\,.
\end{eqnarray}
Identifying the lightest state $H_1$ as the SM-like Higgs boson, one may see
$g_{_{H_1VV}}=O_{\varphi_1 1}=c_\gamma c_\eta$ which should take
the SM value of 1 in the alignment limit. Therefore, in this parameterization,
the alignment limit might be realized by taking $\gamma=\eta=0$ and one may take
the convention of $|\gamma|\leq \pi/2$ and $|\eta|\leq \pi/2$ 
without loss of generality to have positive $g_{_{H_1VV}}$ both in
the CPC ($\eta=0$) and CPV ($\eta\neq 0$) cases.
Note that, even in the alignment limit of $g_{_{H_1VV}}=1$, 
the CP-even state $\varphi_2$ and the CP-odd
one $a$ mix to result in the two heavy states $H_2$  and $H_3$ and the mixing
between $\varphi_2$ and $a$
is described by the CPV mixing angle $\omega$.
Incidentally, by taking $\gamma=\eta=0$ in the alignment limit, we have 
\begin{eqnarray}
\label{eq:Z1456align}
&&
\left.Z_1\right|_{\gamma=\eta=0}=\frac{M_{H_1}^2}{2v^2} \,, \ \ \
\left.Z_4\right|_{\gamma=\eta=0}=\frac{M_{H_2}^2+M_{H_3}^2-2M_{H^\pm}^2}{v^2} \,, \ \ \
\left.Z_6\right|_{\gamma=\eta=0}=0 \,, \nonumber \\[2mm]
&&
\left.Z_5\right|_{\gamma=\eta=0}=
-\frac{M_{H_3}^2-M_{H_2}^2}{2v^2}\,c_{2\omega} \ - \
i\,\frac{M_{H_3}^2-M_{H_2}^2}{2v^2}\,s_{2\omega}\,.
\end{eqnarray}
In the CPC alignment limit of $s_{2\omega}=0$, 
$H_2$ ($H_3$) is CP odd when $|s_\omega|=1$ ($|c_\omega|=1$).
We observe that the upper limits on the absolute values of
the quartic couplings $Z_4$ and $Z_5$ constrain
the product of the mass scale of heavy Higgs bosons and
the mass splitting between them. For example, in the alignment limit,
the upper limit on $|Z_5|$ 
puts the following constraint on 
the mass splitting between the heavy neutral Higgs bosons:
\begin{equation}
\label{eq:Z5delta}
M_{H_3}-M_{H_2} \leq
\frac{2|Z_5|_{\rm max}\,v^2}{M_{H_3}+M_{H_2}}\,,
\end{equation}
which leads to the more degenerate mass spectrum of the two
heavy neutral Higgs bosons as they become heavier.

The Higgs potential contains 3 dimensionful quadratic parameters $Y_{1,2,3}$ and 
7 dimensionless quartic parameters $Z_{1-7}$, see Eq.~(\ref{eq:VHiggs}). 
Using the tadpole conditions given by Eq.~(\ref{eq:higgsbasistadpole}), 
the parameters $Y_1$ and $Y_3$ might be removed in favor of $v$, $Z_1$, and $Z_6$, 
and the remaining dimensionful parameter $Y_2$ can be removed
in favor of $Z_3$ and $M_{H^\pm}$, see Eq.~(\ref{eq:mch}). This observation leads
to the input parameter set
\begin{equation}
\label{eq:inputZ}
{\cal I}_Z=
\left\{v,M_{H^\pm};Z_1,Z_4,Z_5,Z_6;Z_3; Z_2,Z_7\right\}\,.
\end{equation}
Alternatively, one may use the more physical set of input parameters
\begin{equation}
\label{eq:inputP}
{\cal I}_P=
\left\{v,M_{H^\pm};M_{H_1},M_{H_2},M_{H_3},\{O\}_{3\times 3};Z_3; Z_2,Z_7\right\}\,,
\end{equation}
by exploiting the relations given by Eq.~(\ref{eq:Z1456}).
In either of the two sets, there are three parameters involved with CP violation:
$\imag(Z_5)$, $\imag(Z_6)$, and
$\imag(Z_7)$ in ${\cal I}_Z$ or
$\eta$, $\omega$, and $\imag(Z_7)$ in ${\cal I}_P$. 
But one should be cautious to note that 
the three CPV parameters are not all physical
and physical observables depend on only two (combinations) of them.
This could be understood by observing that
the Higgs potential is invariant under the following
phase rotations:
\begin{eqnarray}
\label{eq:rephasing}
{\cal H}_2 \to  {\rm e}^{+i\zeta}\,{\cal H}_2\,; \ 
Z_5 \to Z_5\,{\rm e}^{-2i\zeta}\,, \,
Z_6 \to Z_6\,{\rm e}^{-i\zeta}\,, \,
Z_7 \to Z_7\,{\rm e}^{-i\zeta}\,,
\end{eqnarray}
which lead to the following two rephasing-invariant CPV phases
\begin{eqnarray}
\hspace{-1.0cm}
\theta_1\equiv{\rm Arg}[Z_6 (Z_5^*)^{1/2}] \ \ \ {\rm and} \ \ \
\theta_2\equiv{\rm Arg}[Z_7 (Z_5^*)^{1/2}]\,, 
\end{eqnarray}
pivoting around the complex quartic coupling $Z_5$~\cite{Lee:2021oaj}.

Note that, under phase rotations given by Eq.~(\ref{eq:rephasing}),
the $3\times 3$ mass-squared matrix ${\cal M}_0^2$ of the neutral Higgs bosons
transforms also and, accordingly,
the mixing matrix $O$ is shifted~\cite{Haber:2006ue,Haber:2010bw}.
More precisely, for the convention given 
by~Eqs.(\ref{eq:aOi}) and (\ref{eq:OTM2O}) adopted in this work,
the phase rotation ${\cal H}_2 \to {\rm e}^{i\zeta} {\cal H}_2$ 
or $(\varphi_1,\varphi_2,a)^T \to O_\zeta^T (\varphi_1,\varphi_2,a)^T$
leads to~\cite{Lee:2021oaj}
\begin{equation}
{\cal M}_0^2 \to O_\zeta^T  {\cal M}_0^2 O_\zeta\,; \ \ \
O \ \to \ O_\zeta^T \ O\,,
\end{equation}
with
\begin{equation}
O_\zeta = \left(\begin{array}{ccc}
1 & 0 & 0 \\
0 & c_\zeta & s_\zeta \\
0 & -s_\zeta & c_\zeta
\end{array} \right)\,.
\end{equation}
Exploiting this feature, one may observe that, among
the three mixing angles of $\gamma$, $\eta$, and $\omega$
parameterizing the mixing matrix $O$ as in Eq.~(\ref{eq:omix}), the two CPV angles
$\eta$ and $\omega$ are not always independent.
For example, one may take the input parameter set ${\cal I}_Z$ and choose the
basis in which
$\imag(Z_5)=0$ by an appropriate phase rotation. In this case, one may extract
the three mixing angles $\gamma$, $\eta$, and $\omega$
by diagonalizing the $3\times 3$ mass-squared matrix ${\cal M}_0^2$ 
given by Eq.~(\ref{eq:m0sq}).
But, note that the two CPV angles $\eta$ and $\omega$ cannot be 
independent since
CP violation in the neutral Higgs sector is dictated 
solely by $\imag(Z_6)$  in this case with $\imag(Z_5)=0$.
\footnote{
It is worthwhile to note that this is the direct consequence of
the tree-level analysis of the Higgs potential.
Beyond the tree level,
the $3\times 3$ mass-squared matrix ${\cal M}_0^2$ might receive additional
CPV contributions coming from other than $\imag(Z_5)$ and $\imag(Z_6)$
through higher loop corrections.
In the 2HDM framework, the additional CPV sources include
$\imag(Z_7)$ and $\imag(\zeta_{u,d,e})$ with the latter ones
possibly residing in the Yukawa sector, 
see Eq.~(\ref{eq:yukawa0}) and discussion below.
Regarding the rephasing invariance of all the relevant CPV phases, 
we refer to Ref.~\cite{Lee:2021oaj}.}
In fact, using Eqs.~(\ref{eq:Z1456}) and (\ref{eq:omix}), 
one may see that they are related through the following relation
\begin{equation}
\imag(Z_5)=
\left[ \frac{M_{H_3}^2 c_\omega^2 +M_{H_2}^2 s_\omega^2 -M_{H_1}^2}{v^2}\,
s_\gamma s_\eta
- \frac{M_{H_3}^2-M_{H_2}^2}{v^2}\,
c_\gamma c_\omega s_{\omega}\,\right]\, c_\eta  \ = \  0\,,
\end{equation}
which can be solved for $s_\eta$ when $c_\eta \neq 0$~\cite{Lee:2021oaj}:
\begin{equation}
\label{eq:sineta_IZ50}
\left.s_\eta\right|_{\imag{Z_5}=0} =
\frac{(M_{H_3}^2-M_{H_2}^2)c_\gamma c_\omega s_\omega}
{(M_{H_3}^2 c_\omega^2 +M_{H_2}^2 s_\omega^2 -M_{H_1}^2) s_\gamma}\,.
\end{equation}
Of course, one may choose the basis in which 
both $\imag(Z_5)$ and $\imag(Z_6)$ are non-vanishing.  In this case,
$\eta$ and $\omega$ are independent from each other
though the physical observables such as the masses of the neutral Higgs bosons
and their couplings to two massive vector bosons depend only
on the relative phase $\theta_1={\rm Arg}[Z_6 (Z_5^*)^{1/2}]$
which is invariant under the phase rotations given by
Eq.~(\ref{eq:rephasing}).

\begin{figure}[t!]
\vspace{-0.5cm}
\begin{center}
\includegraphics[width=8.0cm]{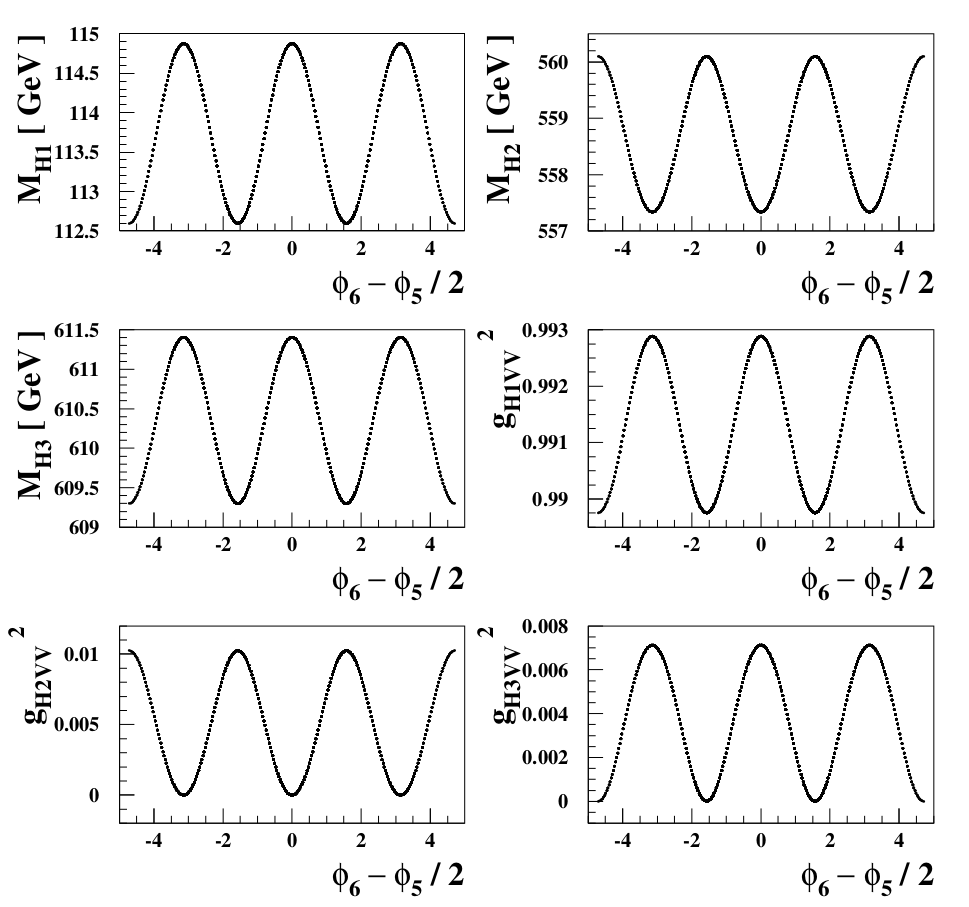}
\includegraphics[width=8.0cm]{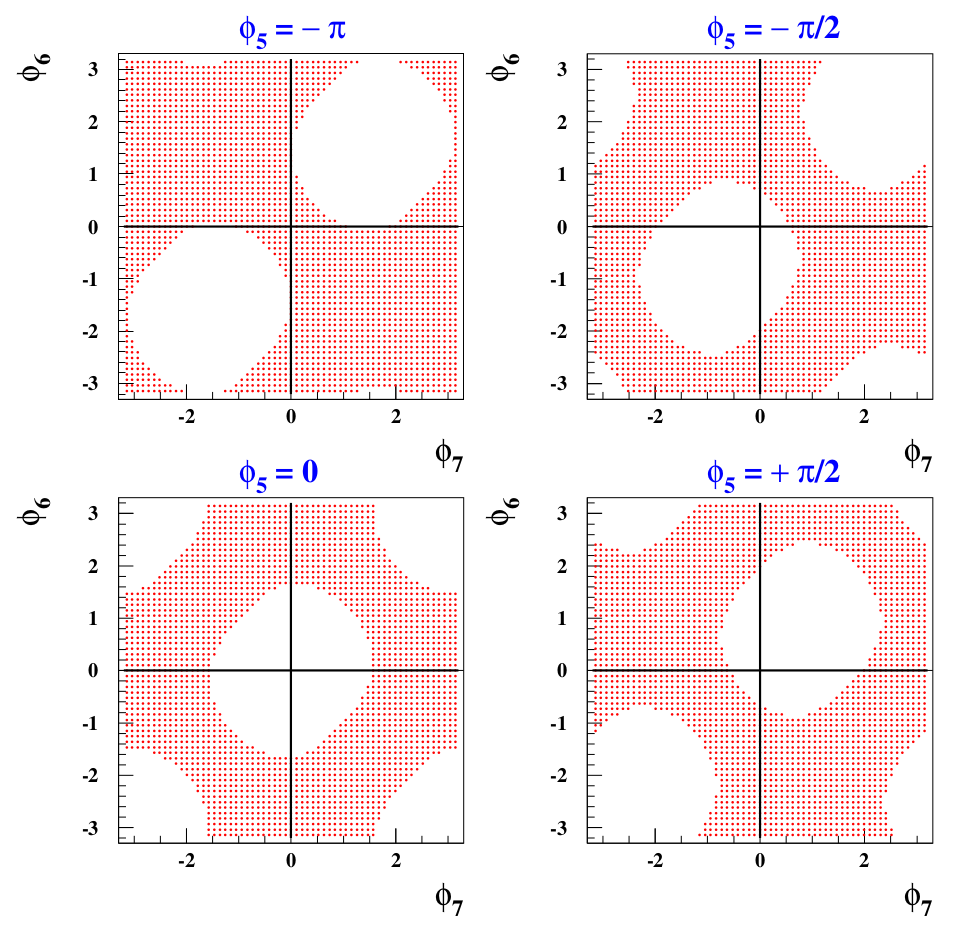}
\end{center}
\vspace{-0.5cm}
\caption{\it 
(Left) The masses of three neutral Higgs
bosons $M_{H_i}$ and their couplings-squared to two massive vector bosons
$g_{{H_iVV}}$ as functions of $\theta_1=\phi_6-\phi_5/2$ taking
$Z_1=0.13$, $Z_3=2.5$, $Z_4=3.0$,
$|Z_5|=|Z_6|=0.5$, and $M_{H^\pm}=500$ GeV
while varying $\phi_5\equiv {\rm Arg}(Z_5)$ and 
$\phi_6 \equiv {\rm Arg}(Z_6)$ from $-\pi$ to $\pi$.
(Right) The allowed region on the $\phi_7$-$\phi_6$ plane
obtained by imposing the UNIT$\oplus$BFB constraints for the
four values of $\phi_5=-\pi$ (upper-left), $-\pi/2$ (upper-right), 
$0$ (lower-left), and $\pi/2$ (lower-right). The quartic couplings
are taken as in the left panel additionally with $Z_2=0.5$ and
$|Z_7|=0.5$.
}
\label{fig:check}
\end{figure}
In the left panel of Fig.~\ref{fig:check}, we show the 
masses of three neutral Higgs
bosons $M_{H_i}$ and their couplings to two massive vector bosons
$g_{{H_iVV}}$ as functions of $\theta_1=\phi_6-\phi_5/2$ taking
$Z_1=0.13$, $Z_3=2.5$, $Z_4=3.0$, 
$|Z_5|=|Z_6|=0.5$, and $M_{H^\pm}=500$ GeV
while varying $\phi_5\equiv {\rm Arg}(Z_5)$ and 
$\phi_6 \equiv {\rm Arg}(Z_6)$ from $-\pi$ to $\pi$.
\footnote{
Note that varying $\phi_5$ and $\phi_6$ separately
is redundant considering the rephasing invariance of the Higgs potential.
But, it serves for our purpose here to check that
the physical observables under consideration indeed depend on the
rephasing-invariant CPV phase of $\theta_1$ only.}
We observe that the physical observables shown depend only on the 
rephasing-invariant CPV phase of $\theta_1$.
Note that $Z_7$ is not involved with the masses and mixing of the
charged and neutral Higgs bosons and, in this case, there is
only one physical CPV phase of $\theta_1$.
On the other hand,
as will be addressed soon in the next section, the perturbative unitarity 
conditions and those for the Higgs potential to be bounded from below
involve all the seven quartic couplings $Z_{1-7}$ and, in this case,
there are two physical CPV phases of 
$\theta_1$ and $\theta_2=\phi_7-\phi_5/2$ with $\phi_7 \equiv {\rm Arg}(Z_7)$.
In the right panel of Fig.~\ref{fig:check}, we show the 
region on the $\phi_7$-$\phi_6$ plane
in which the two conditions are fulfilled taking 
$Z_1=0.13$, $Z_2=0.5$, $Z_3=2.5$, $Z_4=3.0$, and
$|Z_5|=|Z_6|=|Z_7|=0.5$ while varying $\phi_6$ and
$\phi_7$ from $-\pi$ to $\pi$.  For $\phi_5$, we have taken
the four representative values of $-\pi$, $-\pi/2$, $0$, and $\pi/2$.
\footnote{
Note again that varying $\phi_5$, $\phi_6$, and $\phi_7$  separately
is redundant considering the rephasing invariance of the Higgs potential.
But, it serves for our purpose here to check that
the physical conditions under consideration indeed depend only on the
two rephasing-invariant CPV phases of $\theta_1$ and $\theta_2$.}
We observe that the allowed regions shift along the $\phi_6=\phi_7$ line
by the amount of $\phi_5/2$ as they should do.
Incidentally, we also explicitly check the relation 
Eq.~(\ref{eq:sineta_IZ50})
in the subset of the parameter space 
in which $\imag(Z_5)=0$ and $c_\eta \neq 0$ by choosing the 
input parameter set ${\cal I}_Z$ while scanning
all the three CPV potential parameters of
$\imag(Z_5)$, $\imag(Z_6)$, and $\imag(Z_7)$ redundantly.

For our study, we choose the set  ${\cal I}_P$ given by Eq.~(\ref{eq:inputP})
for the input  parameters:
$$
{\cal I}_P=
\left\{v,M_{H^\pm};M_{H_1},M_{H_2},M_{H_3},\{O\}_{3\times 3};Z_3; Z_2,Z_7\right\}\,,
$$
with $\{O\}_{3\times 3}=\gamma\,,\eta\,,\omega$.
The set ${\cal I}_P$ contains 12 real degrees of freedom including $v$.
For the angles $\gamma$ and $\eta$, we take
the convention of $|\gamma|\leq \pi/2$ and $|\eta|\leq \pi/2$ 
without loss of generality resulting in
$c_\gamma \geq 0$ and $c_\eta \geq 0$ and the angle $\omega$ is varied between
$-\pi$ to $\pi$.
We identify lightest neutral Higgs boson $H_1$
as the SM-like one with $M_{H_1}=125.5$ GeV and
the heavy Higgs masses squared are scanned up to $(1.5~{\rm TeV})^2$.
Finally, the quartic couplings $Z_{2}$, $|Z_{3}|$, $|\real(Z_7)|$, and $|\imag(Z_7)|$
are scanned up to 3, 10, 5, respectively.
\footnote{Note that $Z_2\geq 0$, see Eq.(\ref{eq:bfb}).}
Though
our choice of parameters is redundant considering  the rephasing invariance of
the Higgs potential under the phase rotations given by Eq.~(\ref{eq:rephasing}),
it should be useful for checking the rephasing invariance and
the consistency of our numerical analysis explicitly.

\section{Constraints}
\label{sec:constraints}
First, we consider the  perturbative unitarity (UNIT) conditions and
those for the Higgs potential to be bounded from below (BFB) to obtain the
primary theoretical constraints on the potential parameters 
or, equivalently, the constraints on the Higgs-boson masses 
and the three mixing angles including 
correlations among them.
The unitarity conditions in the  most general 2HDM have been investigated 
in Refs.~\cite{Ginzburg:2005dt,Kanemura:2015ska,Jurciukonis:2018skr}.
On the other hand,
the necessary and sufficient conditions for the 2HDM potential to
be bounded from below have been derived in
Refs.~\cite{Maniatis:2006fs,Ivanov:2006yq,Ivanov:2007de,Ivanov:2015nea}.
Recently, the authors of Ref.~\cite{Bahl:2022lio}
have provided some analytic expressions for the UNIT and BFB conditions
in the  most general 2HDM.
In this work, 
for the unitarity conditions, we closely follow Ref.~\cite{Jurciukonis:2018skr}
taking into account three scattering matrices
which are expressed in terms of the quartic couplings $Z_{1-7}$.
Using the set ${\cal I}_P$ in Eq.~(\ref{eq:inputP})
for the input parameters,
all the seven quartic couplings are fixed exploiting the relations
in Eq.~(\ref{eq:Z1456}). 
For the details of the implementation of the UNIT conditions, we refer to 
Refs.~\cite{Jurciukonis:2018skr,Lee:2021oaj}.
For the BFB constraints, we require the following 5 necessary conditions
for the Higgs potential to be bounded-from-below 
\cite{Branco:2011iw}:
\begin{eqnarray}
\label{eq:bfb}
Z_1 \geq 0 \,, \ \ \ Z_2 \geq 0\,;&& \nonumber \\[2mm]
2\sqrt{Z_1Z_2}+Z_3 \geq  0 \,, \ \ \  2\sqrt{Z_1Z_2}+Z_3+Z_4-2|Z_5|\geq 0\,;&& \nonumber \\[2mm]
Z_1+Z_2+Z_3+Z_4+2|Z_5|-2|Z_6+Z_7|  \geq 0\,.&&
\end{eqnarray}
Note that the first 4 inequalities are necessary and sufficient conditions
when $Z_6=Z_7=0$.  Otherwise, they constitute only the necessary conditions
together with the last one. 
The authors of Ref.~\cite{Jurciukonis:2018skr} numerically analyze the
BFB conditions when $Z_6$ and $Z_7$ are nonzero and complex 
and confirm that the five inequalities always hold at tree level.
We have explicitly checked that the parameter space where
the five inequalities hold fully contains the parameter space in which
the necessary and sufficient BFB conditions are fulfilled.

Second, we consider
the electroweak (ELW) oblique corrections to the so-called $S$, $T$ and $U$
parameters~\cite{Peskin:1990zt,Peskin:1991sw} which provide significant
constraints on the quartic couplings of the 2HDM.
Fixing $U=0$ which is suppressed by
an additional factor $M_Z^2/M^2_{\rm BSM}$
\footnote{Here, $M_{\rm BSM}$ denotes some heavy mass scale involved
with new physics beyond the Standard Model (BSM).} relative to 
$S$ and $T$,
the $S$ and $T$ parameters are constrained as follows
\cite{ParticleDataGroup:2020ssz,Lee:2012jn}
\begin{equation}
\label{eq:STRange}
\frac{(S-\widehat S_0)^2}{\sigma_S^2}\ +\
\frac{(T-\widehat T_0)^2}{\sigma_T^2}\ -\
2\rho_{ST}\frac{(S-\widehat S_0)(T-\widehat T_0)}{\sigma_S \sigma_T}\
\leq\ R^2\,(1-\rho_{ST}^2)\; ,
\end{equation}
with $R^2=2.3$, $4,61$, $5.99$, $9.21$, $11.83$ at $68.3 \%$, $90 \%$,
$95 \%$, $99 \%$, and $99.7 \%$  confidence levels (CLs), respectively.
For our numerical analysis, we take the 95\% CL limit. 
For the central values $\widehat S_0$ and $\widehat T_0$,
the standard deviations $\sigma_{S,T}$, and the correlations between them,
we adopt~\cite{ParticleDataGroup:2020ssz,Lu:2022bgw}:
\begin{eqnarray}
\label{eq:STEXP}
&{\bf PDG}&~:~~~
(\widehat S_0\,,\ \sigma_S)\ =\ (0.00\,,\ 0.07)\;,\qquad
(\widehat T_0\,,\ \sigma_T)\ =\ (0.05\,,\ 0.06)\;,\qquad
\rho_{ST}=0.92\,, \nonumber \\
&{\bf CDF}&~:~~~
(\widehat S_0\,,\ \sigma_S)\ =\ (0.15\,,\ 0.08)\;,\qquad
(\widehat T_0\,,\ \sigma_T)\ =\ (0.27\,,\ 0.06)\;,\qquad
\rho_{ST}=0.93\,.
\end{eqnarray}

Using the set ${\cal I}_P$ for the input parameters and exploiting
the relations among the couplings of the Higgs bosons to the 
massive vector bosons given in Eq.~(\ref{eq:2hdmsumrule}),
the $S$ and $T$ parameters might be estimated using the following expressions
at the one-loop order
\cite{Toussaint:1978zm,Grimus:2007if,Grimus:2008nb,
Haber:2010bw,Branco:2011iw,Lee:2012jn,Lee:2021oaj}:
\footnote{To obtain the expressions for the $T$ and $S$ parameter,  
we use the coupling relations  
$g_{_{H_iH_jV}}^2=|\epsilon_{ijk}|\,g_{_{H_kVV}}^2$ and
the sum rule 
$|g_{_{H_iH^+ W^-}}|^2\ =\ 1 - g_{_{H_iVV}}^2$ for $i,j=1,2,3$,
see Eqs.~(\ref{eq:2hdmhvvetc}) and (\ref{eq:2hdmsumrule}).}
\begin{eqnarray}
\label{eq:ST}
\frac{T}{\frac{\sqrt{2}G_F}{16\pi^2\alpha_{\rm EM}}}
&=&
 (1-g_{_{H_1VV}}^2)F_\Delta(M_{H_1},M_{H^\pm})
+(1-g_{_{H_2VV}}^2)F_\Delta(M_{H_2},M_{H^\pm})
+(1-g_{_{H_3VV}}^2)F_\Delta(M_{H_3},M_{H^\pm})
\nonumber \\ 
&&
-g_{_{H_3VV}}^2 F_\Delta(M_{H_1},M_{H_2})
-g_{_{H_2VV}}^2 F_\Delta(M_{H_1},M_{H_3})
-g_{_{H_1VV}}^2 F_\Delta(M_{H_2},M_{H_3})\,,
\nonumber \\[2mm]
4\pi\,S &=&
-F^\prime_\Delta(M_{H^\pm},M_{H^\pm})
+g_{_{H_3VV}}^2 F^\prime_\Delta(M_{H_1},M_{H_2})
+g_{_{H_2VV}}^2 F^\prime_\Delta(M_{H_1},M_{H_3})
+g_{_{H_1VV}}^2 F^\prime_\Delta(M_{H_2},M_{H_3})\,,
\end{eqnarray}
where the one-loop functions are given by
\begin{eqnarray}
\label{eq:ffp}
F_\Delta(m_0,m_1) &=& F_\Delta(m_1,m_0) =
\frac{m_0^2+m_1^2}2 -\frac{m_0^2m_1^2}{m_0^2-m_1^2}\ln\frac{m_0^2}{m_1^2}\,,
\nonumber \\[3mm]
F_\Delta^\prime(m_0,m_1) &=& F_\Delta^\prime(m_1,m_0) =
-\frac{1}{3} \left[ \frac{4}{3}
-\frac{m_0^2 \ln m_0^2 -m_1^2 \ln m_1^2}{m_0^2-m_1^2}
-\frac{m_0^2+m_1^2}{(m_0^2-m_1^2)^2}F_\Delta(m_0,m_1) \right]\,.
\end{eqnarray}
\begin{figure}[t!]
\vspace{-0.5cm}
\begin{center}
\includegraphics[width=12.0cm]{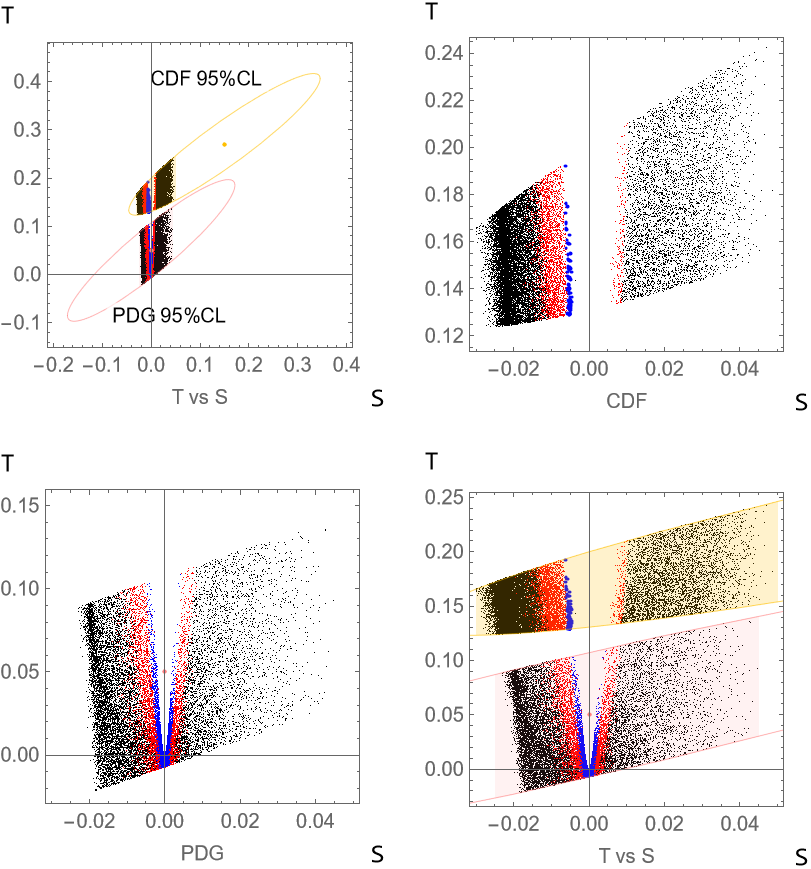}
\end{center}
\vspace{-0.5cm}
\caption{\it  {\bf CDF} and {\bf PDG}:
$T$ versus $S$. 
The scattered regions are obtained by imposing the combined
UNIT$\oplus$BFB$\oplus$ELW$_{95\%}$
constraints and the colored dots are for 
$M_{H^\pm}>M_{H_1}$ (black$+$red$+$blue),
$M_{H^\pm}>$ 500 GeV (red$+$blue), and
$M_{H^\pm}>$ 900 GeV (blue).
}
\label{fig:TversusS}
\end{figure}

When the BSM Higgs bosons are almost degenerate
and they are much heavier than the SM-like one,
it is useful to know the asymptotic behavior
of the one-loop functions.
When $m_0\sim m_1$, we find 
\footnote{
Note that the one-loop functions $F_\Delta(m_0,m_1)$ and 
$F^\prime_\Delta(m_0,m_1)$ are symmetric 
under the exchange $m_0\leftrightarrow m_1$.}
\begin{eqnarray}
\label{eq:fapp1}
F_\Delta(m_0,m_1) &=& \frac{2(m_0-m_1)^2}{3} 
-\frac{2\,(m_0-m_1)^4}{15\,(m_0+m_1)^2}
+{\cal O}\left[\frac{(m_0-m_1)^6}{(m_0+m_1)^4}\right]\,, \nonumber \\[2mm]
F^\prime_\Delta(m_0,m_1) &=& 
\frac{1}{3}\ln\left[\frac{(m_0+m_1)^2}{4}\right]
+ \frac{(m_0-m_1)^2}{5\,(m_0+m_1)^2}
+{\cal O}\left[\frac{(m_0-m_1)^4}{(m_0+m_1)^4}\right]\,.
\end{eqnarray}
On the other hand, for $m_1\gg m_0$, we have
\begin{eqnarray}
\label{eq:fapp2}
F_\Delta(m_0,m_1) &=& \frac{m_1^2}{2}
+\left(\frac{1}{2}+\ln\frac{m_0^2}{m_1^2}\right)\,m_0^2
+{\cal O}\left[\left(\frac{m_0^4}{m_1^2}\right)
\ln\frac{m_0^2}{m_1^2}\right]\,, \nonumber \\[2mm]
F^\prime_\Delta(m_0,m_1) &=& \frac{\ln m_1^2}{3}
-\frac{5}{18} +\frac{2}{3}\frac{m_0^2}{m_1^2}
+{\cal O}\left[\left(\frac{m_0^4}{m_1^4}\right)
\ln\frac{m_0^2}{m_1^2}\right]\,.
\end{eqnarray}
In the alignment limit where 
$g_{_{H_1VV}}^2=1$ and  $g_{_{H_2VV}}^2=g_{_{H_3VV}}^2=0$, we note that
$S$ and $T$ are symmetric under the exchange $M_{H_2}\leftrightarrow M_{H_3}$
and they are identically vanishing if $M_{H_2}=M_{H_3}=M_{H^\pm}$ since
$F_\Delta(m,m)=0$ and
$F^\prime_\Delta(m,m)=\ln m^2/3$.
In the upper-left panel of Fig.~\ref{fig:TversusS}, we show the two ELW ellipses 
which delimit the CDF (upper ellipse) and PDG (lower ellipse) 95\% CL regions 
on the $S$-$T$ plane, see Eqs.~(\ref{eq:STRange}) and (\ref{eq:STEXP}).
The scattered region in each of them has been obtained by
imposing the UNIT and {\it necessary}
BFB constraints as well as the ELW one
as indicated by the abbreviation UNIT$\oplus$BFB$\oplus$ELW$_{95\%}$ in the figure caption.
We magnify the CDF and PDG regions in the upper-right and lower-left panels, respectively,
with the colored dots for 
$M_{H^\pm}>M_{H_1}$ (black$+$red$+$blue),
$M_{H^\pm}>$ 500 GeV (red$+$blue), and
$M_{H^\pm}>$ 900 GeV (blue).
We observe that $M_{H^\pm}$ is smaller than 900 GeV for CDF when $S$ is positive, 
see the upper-right panel.
In the lower-right plane, we show the two magnified CDF and PDG regions together
for simultaneous comparisons.
We find that $S$ takes values in the range between $-0.03$ and $0.05$ whose
absolute values are smaller than $\sigma_S=0.07$ (PDG) and 
$0.08$ (CDF), see Eq.~(\ref{eq:STEXP}).
The $T$ parameter takes its value between $0.12$ and $0.24$ (CDF)
and between $-0.02$ and $0.14$ (PDG).
Note that, for CDF, the narrow region around $S=0$ with radius about $0.005$ 
is not allowed, see the upper-right panel. Also note that $T$ is 
positive definite and sizable.

\begin{figure}[t!]
\vspace{-0.5cm}
\begin{center}
\includegraphics[width=8.0cm]{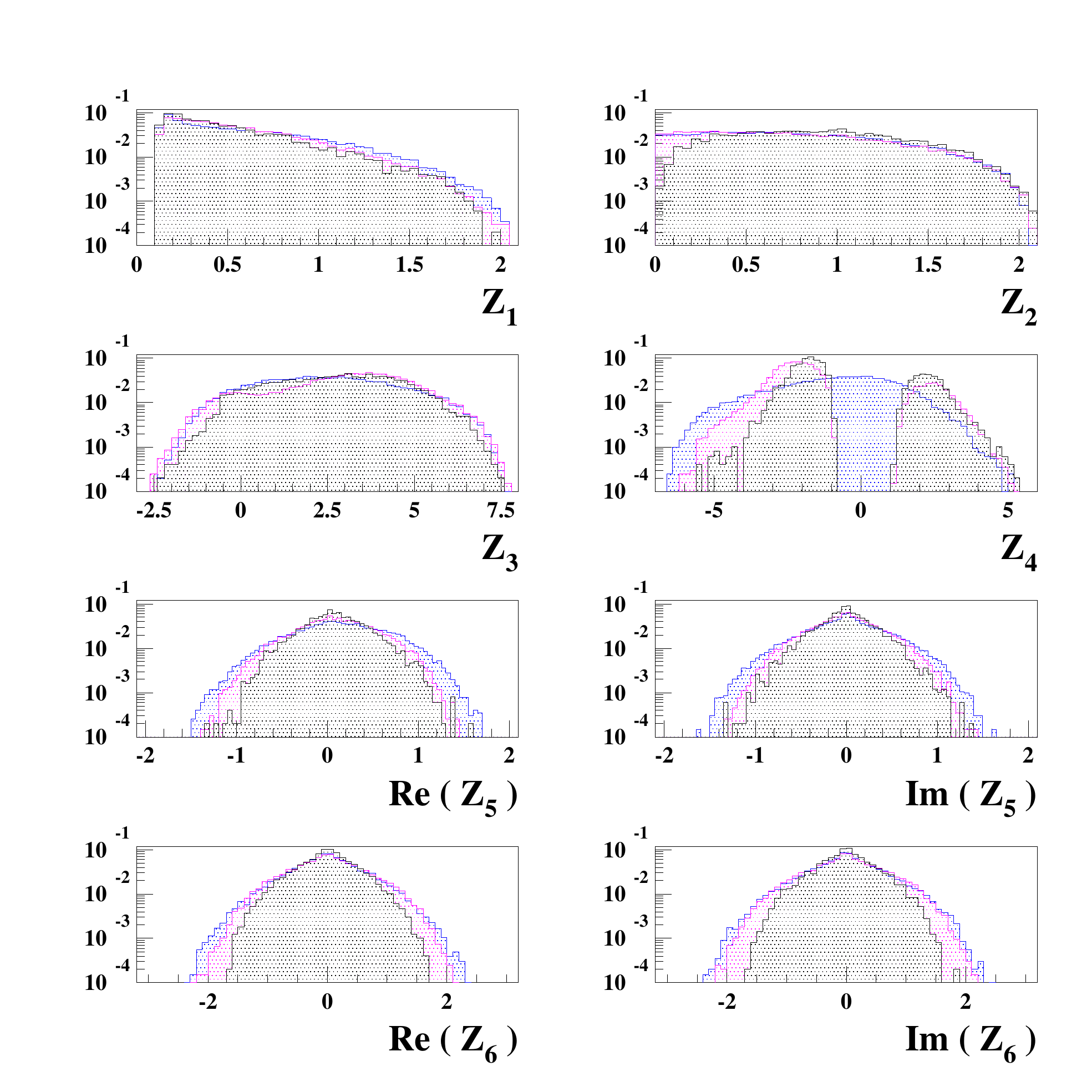}
\includegraphics[width=8.0cm]{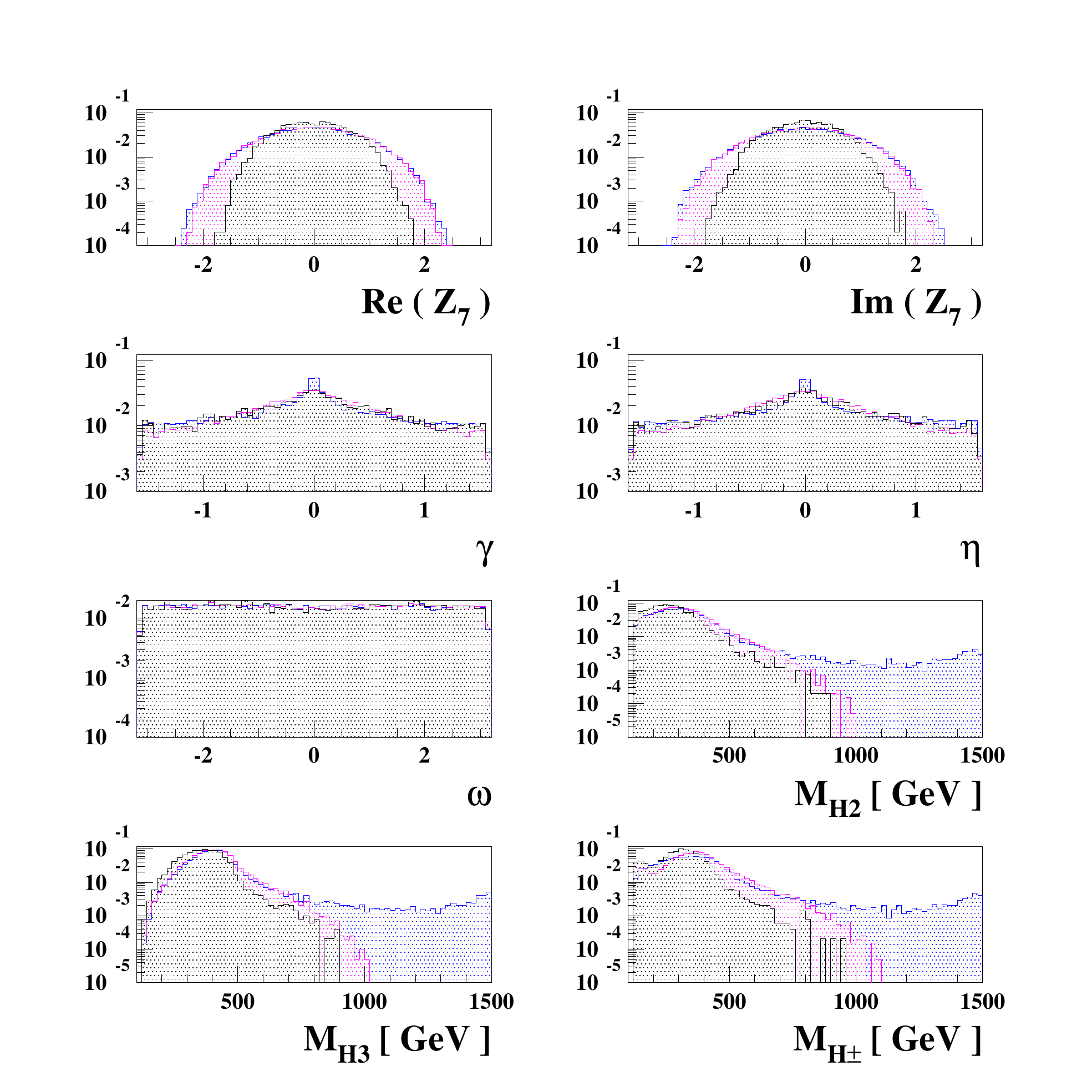}
\end{center}
\vspace{-0.5cm}
\caption{\it {\bf CDF}:
The normalized distributions of the quartic couplings,
the mixing angles, and the masses of heavy Higgs bosons
imposing the combined UNIT$\oplus$BFB$\oplus$ELW$_{95\%}$ 
constraints (magenta) and using the set ${\cal I}_P$ of the input parameters, 
see Eq.~(\ref{eq:inputP}).
The normalized distributions obtained by replacing the necessary BFB conditions
with the necessary and sufficient ones are shown in black.
For comparisons, we also show the results obtained by imposing
only the UNIT and necessary BFB constraints (blue).
}
\label{fig:ziang}
\end{figure}
In Fig.~\ref{fig:ziang}, we show the normalized distributions (magenta)
of the seven quartic couplings,
the three mixing angles, and the masses of heavy Higgs bosons 
using the CDF values for 
$\widehat{T}_0$, $\widehat{S}_0$, $\sigma_S$, $\sigma_T$, and $\rho_{ST}$,
see Eq.~(\ref{eq:STEXP}).
The combined UNIT$\oplus$BFB$\oplus$ELW$_{95\%}$
constraints have been imposed. Also shown are the distributions (blue)
requiring only the UNIT$\oplus$BFB conditions.
We find that the distributions of the real and imaginary parts of $Z_{5,6,7}$ are 
(almost) symmetric around 0.
Meanwhile, the $Z_4$ distribution changes drastically
to exclude the region $|Z_4|\lsim 1$
after imposing the ELW constraint additionally.
Quantitatively, we find that the quartic couplings are constrained as:
\begin{eqnarray}
&& 
0.1\lsim Z_1 \lsim 2.0\,, \ \ \
0  \lsim Z_2 \lsim 2.1\,, \ \ \
-2.6 \lsim Z_3 \lsim 8.0\,, \nonumber \\[2mm]
&& -6.3 \lsim Z_4 \lsim -0.8 \ \cup \
1.1 \lsim Z_4 \lsim 5.4\,,
\nonumber \\[2mm]
&&
|Z_5|\lsim 1.7\,, \ \ \
|Z_6|\lsim 2.4\,, \ \ \
|Z_7|\lsim 2.7\,. \ \ \
\end{eqnarray}
The mixing angles $\gamma$ and $\eta$
prefer the values near to 0 while the $\omega$ distribution
is flat.
From the the distributions of the masses of heavy Higgs bosons, we
observe that the upper limit around 1 TeV emerges
when the ELW constraint has been imposed in addition to 
the UNIT$\oplus$BFB ones.
In Fig.~\ref{fig:ziang}, we also show the normalized distributions 
(black) by picking out the points which satisfy the necessary and
sufficient BFB conditions as well as the UNIT$\oplus$ELW$_{95\%}$ constraints.
We observe that the stronger BFB conditions 
shift the negative edge of $Z_4$ to the positive 
direction a little bit and 
reduce the ranges of $Z_6$ and $Z_7$ couplings 
as $|Z_{6,7}| \lsim 1.9$ and 
strengthen the upper limit on $M_{H^\pm,H_{2,3}}$
by the amount of about 100 GeV.
Otherwise, the distributions remain the same more or less 
especially with the gap around $Z_4=0$ unaffected.

\begin{figure}[t!]
\vspace{-0.5cm}
\begin{center}
\includegraphics[width=14.0cm]{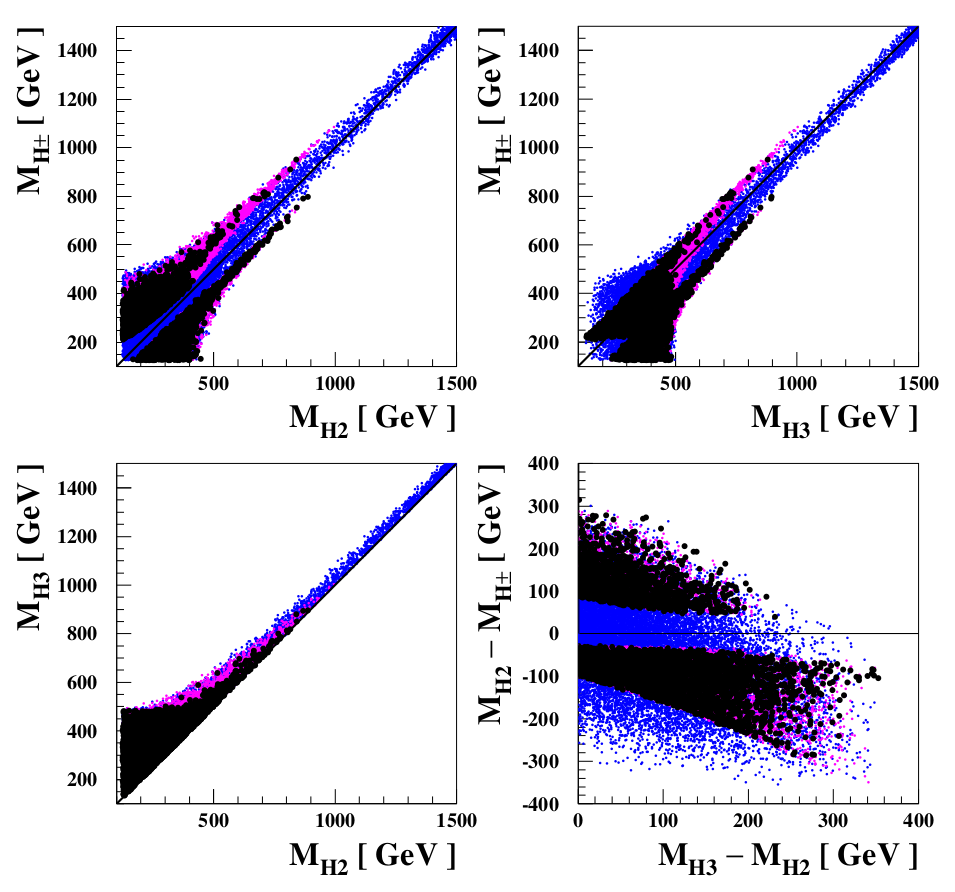}
\end{center}
\vspace{-0.5cm}
\caption{\it {\bf CDF}:
Scatter plots of 
$M_{H^\pm}$ versus $M_{H_2}$ (upper left),
$M_{H^\pm}$ versus $M_{H_3}$ (upper right),
$M_{H_3}$ versus $M_{H_2}$ (lower left), and
$M_{H_2}-M_{H^\pm}$ versus $M_{H_3}-M_{H_2}$ (lower right)
imposing the combined UNIT$\oplus$BFB$\oplus$ELW$_{95\%}$ 
constraints (magenta) and using the set ${\cal I}_P$ of the input parameters, 
see Eq.~(\ref{eq:inputP}).
Scatter plots obtained by replacing the necessary BFB conditions
with the necessary and sufficient ones are shown in black.
For comparisons, we also show the results obtained by imposing
only the UNIT and necessary BFB constraints (blue).
}
\label{fig:mmcorr}
\end{figure}
In Fig.~\ref{fig:mmcorr}, we show the correlations between
the heavy Higgs masses
$M_{H^\pm}$ and $M_{H_2}$ (upper left),
$M_{H^\pm}$ and $M_{H_3}$ (upper right), and
$M_{H_3}$ and $M_{H_2}$ (lower left), and 
that between the mass differences
$M_{H_2}-M_{H^\pm}$ and $M_{H_3}-M_{H_2}$ (lower right)
using the CDF data, see Eq.~(\ref{eq:STEXP}).
As in Fig.~\ref{fig:ziang}, the combined UNIT$\oplus$BFB$\oplus$ELW$_{95\%}$
constraints have been imposed (magenta) and the results 
obtained by adopting the necessary and sufficient BFB conditions
are shown in black.
The blue points are again for the case only with
the UNIT and necessary BFB constraints.
In the upper-left and lower-right panels,
it is clearly shown that there should be sizable mass difference 
between the charged Higgs boson $H^\pm$ and the second heaviest neutral one $H_2$
to accommodate the CDF $W$-mass anomaly.
Further we observe that,
the upper limits of about $1$ TeV and $900$ GeV
on the heaviest states 
emerge when $M_{H^\pm}>M_{H_2}$ and $M_{H^\pm}<M_{H_2}$, respectively.
Note  that, when $M_{H^\pm}<M_{H_2}$, the mass splitting between
$H^\pm$ and $H_2$ also implies the mass splitting between
$H^\pm$ and $H_3$ since $M_{H_2}<M_{H_3}$ by definition.
On the other hand, when $M_{H^\pm}>M_{H_2}$, the heaviest 
neutral Higgs boson could be
either heavier or lighter than the charged Higgs boson, see the upper-right
panel of Fig.~\ref{fig:mmcorr}.
Incidentally, we observe the mass splitting between the two heavy neutral Higgs
bosons limited to be smaller than about 350 (240) GeV when
$M_{H^\pm}>(<)M_{H_2}$, see the lower-right
panel of Fig.~\ref{fig:mmcorr}.
When the masses of the heavy Higgs bosons saturate their upper limits,
there are two types of the heavy-Higgs mass spectrum given by
\begin{eqnarray}
\label{eq:mnc}
M_{H^\pm} > M_{H_2}\,:\, \
&&
M_{H_2}\simeq M_{H_3} \simeq 1000\,(900)\,{\rm GeV}\,, \ \ \ 
M_{H^\pm}\simeq 1100\,(1000)\,{\rm GeV}\,; \ \ \
\nonumber \\[2mm]
M_{H^\pm} < M_{H_2}\,:\, \
&&
M_{H_2} \simeq M_{H_3}\simeq 900\,{\rm GeV}\,, 
\hspace{1.4cm}
M_{H^\pm}\simeq 800\,{\rm GeV}\,,
\end{eqnarray}
where the numbers in parentheses when $M_{H^\pm} > M_{H_2}$
are for the case
with the necessary and sufficient BFB conditions,
see the black points in Fig.~\ref{fig:mmcorr}.
When $M_{H^\pm} < M_{H_2}$, we observe 
that the mass spectrum of heavy Higgs bosons is insensitive to
the choice of the BFB conditions.

In this work, we analyze the mass spectrum of heavy Higgs bosons 
appearing in the CPV 2HDM especially when the central value 
of the $T$ parameter significantly
deviates from its SM value of zero
as indicated by the CDF $W$-mass anomaly
by combining it with the theoretical UNIT and BFB constraints.
In this case, all the relevant parameters and couplings are
fully determined by the CPV 2HDM potential only.
Note that the theoretical UNIT and BFB conditions are given
in terms of the seven quartic couplings $Z_{1-7}$ and
the $S$ and $T$ parameters are functions of the Higgs masses
and the neutral Higgs couplings to a pair of 
massive vector bosons $g_{_{H_iVV}}=O_{\varphi_1 i}$ only, 
see Eq.~(\ref{eq:ST}).
This makes our current analysis largely independent of the Yukawa sector.

Eventually, of course, one needs to implement
a full phenomenological study of the Higgs bosons 
which incorporates 
flavor observables, the LHC Higgs precision data, 
the collider limits from the heavy Higgs boson searches 
carried out at the LHC, non-observation of 
the electron and  neutron electric dipole moments (EDMs), etc.
But, to incorporate all the phenomenological constraints,
one should specify the Yukawa sector since the constraints 
beyond the UNIT, BFB, and ELW ones
strongly depend on the relevant Yukawa interactions.
For example, the recent analysis of the radiative $b\to s\gamma$ decay 
within the so-called type-II 2HDM yields the 95\% CL constraint of
$M_{H^\pm}>800$ GeV~\cite{Misiak:2020vlo}. But it is well known that
the strong upper limit  on the charged Higgs boson mass becomes significantly
weaker if different models for the Yukawa interactions are assumed.
Indeed, there are wider varieties in choosing the 2HDM Yukawa sector
beyond the frequently mentioned four types of 2HDM. 
In the 2HDM, the Yukawa couplings might be given by~\cite{Pich:2009sp,Lee:2021oaj}
\footnote{Here $Q_L^0=(u_L^0\,, d_L^0)^T$, $L_L^0=(\nu_L^0\,, e_L^0)^T$,
$u_R^0$, $d_R^0$, and $e_R^0$ denote the electroweak eigenstates and
$\widetilde{\cal H}_i = i\tau_2 {\cal H}_i^*$.}
\begin{equation}
\label{eq:yukawa0}
-{\cal L}_Y=\sum_{k=1,2}\,
\overline{Q_L^0}\,{\bf y}^u_k\,\widetilde{\cal H}_k\,u_R^0 \ + \
\overline{Q_L^0}\,{\bf y}^d_k\,{\cal H}_k\,d_R^0 \ + \
\overline{L_L^0}\,{\bf y}^e_k\,{\cal H}_k\,e_R^0 \ + \ {\rm h.c.}\,,
\end{equation}
which contain the six $3\times 3$ Yukawa matrices  ${\bf y}_{1,2}^{u,d,e}$.
The Yukawa matrices of ${\bf y}_{1}^{u,d,e}$ are diagonalized to
generate the masses of the SM fermions in the Higgs basis, 
while the Yukawa matrices of 
${\bf y}_{2}^{u,d,e}$ give rise to unwanted 
tree-level Higgs-mediated flavor-changing neutral currents (FCNCs).
To avoid the tree-level FCNC, the Glashow-Weinberg condition
\cite{Glashow:1976nt} has been usually considered leading to 
the conventional four types of 2HDM. But this is not the only way to avoid
the FCNC in the 2HDM. 
For example, one may assume the alignment 
of the two types of the Yukawa matrices or 
${\bf y}_{2}^f=\zeta_f {\bf y}_{1}^f$ 
for $f=u,d,e$
by introducing
the three complex alignment parameters $\zeta_{u,d,e}$~\cite{Pich:2009sp}.
The aligned 2HDM (A2HDM) accommodates the four conventional types of
2HDM as some limiting cases
when the alignment parameters are real and fully 
correlated~\cite{Pich:2009sp,Lee:2021oaj}.
\footnote{For example, the so-called type-II 2HDM can be accommodated
by taking $\zeta_d=\zeta_e=-1/\zeta_u$ and 
$\zeta_u=1/\tan\beta$.}
Otherwise,
it provides the more general 2HDM framework than the conventional 
ones based on the Glashow-Weinberg condition,
leading to much more enriched Higgs phenomenology.
For example,  a very recent study shows that some LHC excesses 
observed by the ATLAS and CMS collaborations
in their searches for heavy neutral scalars
with $\tau^+\tau^-$ and $t\bar t$ final states
are incompatible within the four conventional 2HDMs
but can be accommodated within A2HDM~\cite{Connell:2023jqq}.
This should imply that one needs to be very careful 
when interpreting experimental data if 
there exist several theoretical frameworks available
in order not to exclude regions of parameter space 
which could be important.
A full phenomenological study of the heavy Higgs bosons including
a detailed analysis of various constraints
in the CPV-A2HDM framework
might deserve another independent studies and,
in this work, we keep our analysis independent of the 
Yukawa sector as much as possible.

\section{Analysis}
\label{sec:analysis}
In this section, we provide the comprehensive analysis on the behavior of the masses of
heavy Higgs bosons shown in Fig.~\ref{fig:mmcorr} and investigate 
their dependence on the couplings of the neutral
Higgs bosons to two massive vector bosons $g_{_{H_iVV}}$ 
appearing in the expressions for the $S$ and $T$ parameters, 
see Eq.~(\ref{eq:ST}), taking account of the mass difference $M_{H_3}-M_{H_2}$.
For our analysis, we adopt the necessary BFB conditions 
as in Eq.~(\ref{eq:bfb}) or the magenta points in Fig.~\ref{fig:mmcorr} 
for the conservative estimation of 
the mass spectrum of heavy Higgs bosons taking account of
the limited size of our data set
and also
for the easier comparisons with our previous work~\cite{Heo:2022dey}.
Otherwise, it would be  understood that one might have 
the slightly lighter spectrum
by the amount of about 100 GeV when the charged Higgs boson
is the heaviest state.

The coupling of the SM-like lightest Higgs boson to two massive vector bosons,
$g_{_{H_1VV}}$, is constrained by
the precision LHC Higgs data. Denoting
\begin{equation}
g_{_{H_1VV}} \equiv 1-\epsilon\,,
\end{equation}
the quantity $\epsilon$ is 
required to be smaller than about $0.05$ at $1\sigma$ level~\cite{Cheung:2018ave}.
\footnote{
In Ref.~\cite{Cheung:2018ave},
used are the
accumulated LHC Higgs data with the
integrated luminosities per experiment of approximately
5/fb at 7 TeV, 20/fb at 8 TeV, and up to 80/fb at 13 TeV.
We note that there are more datasets at 13 TeV up to 139/fb and 137/fb
collected with the ATLAS and CMS experiments, respectively,
see Refs.~\cite{ATLAS:2020qdt,CMS:2020gsy}.
We observe that
the $1\sigma$ errors are reduced by the amount of about 30\% by comparing the
results presented in Ref.~\cite{ATLAS:2020qdt} with those in
Ref.~\cite{Aad:2019mbh} in the latter the dataset up to 80/fb is used.
But, without a combined ATLAS and CMS analysis,
it is difficult to say conclusively
how much the full 13-TeV dataset improves the measurements involved with 
the 125 GeV Higgs boson
and, throughout this work, we use $g_{_{H_1VV}}\gsim 0.95$ as
the conservative constraint on the coupling at $1\sigma$ level.}
As a measure of the deviation from the alignment limit
of $g_{_{H_1VV}}^2=1$, we introduce the average coupling
\begin{equation}
g_{23}^2\equiv \frac{g_{_{H_2VV}}^2+g_{_{H_3VV}}^2}{2} 
= \frac{1-g_{_{H_1VV}}^2}{2} 
=\epsilon-\frac{\epsilon^2}{2}\,,
\end{equation}
which varies between $0$ and $0.5$ while being
constrained to be smaller than about $0.05$ at $1\sigma$ level
by the precision LHC Higgs data.

To analyze the CDF case presented in Fig.~\ref{fig:mmcorr},
we introduce the following two measures for the mass splittings of the heavy Higgs bosons:
\begin{equation}
\Delta \equiv M_{H_2}-M_{H^\pm}\,; \ \
\delta \equiv M_{H_3}-M_{H_2}\,,
\end{equation}
and then the masses of heavy Higgs bosons are given by
\begin{equation}
M_{H^\pm}=M-\Delta/2\,, \ \ \
M_{H_2}=M+\Delta/2\,, \ \ \
M_{H_3} = M+\Delta/2+\delta\,,
\end{equation}
in terms of $\Delta$, $\delta$, and the average heavy mass scale
$M=(M_{H^\pm}+M_{H_2})/2$.
The quantity $\delta$ is positive definite while $\Delta$ can be either positive or negative.
When $\Delta<0$, the lightest state is $H_2$ while
$H^\pm$ is the lightest one when $\Delta>0$.

\begin{figure}[t!]
\vspace{-0.5cm}
\begin{center}
\includegraphics[width=8.0cm]{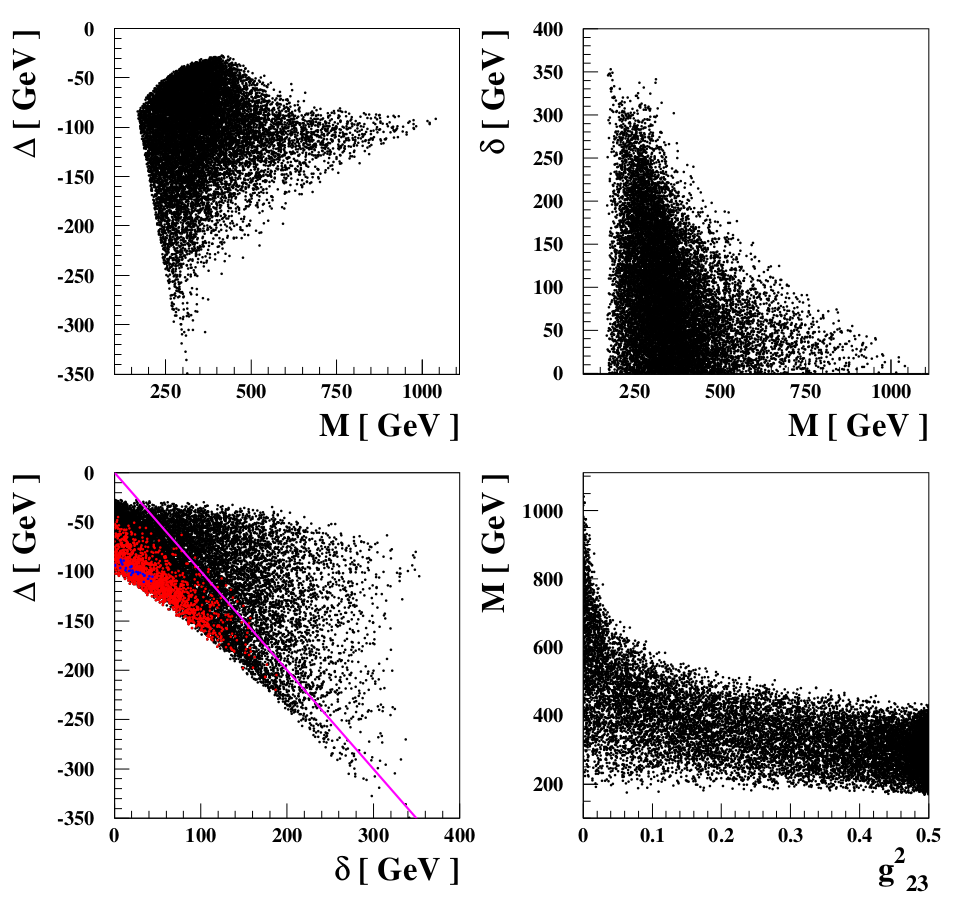}
\includegraphics[width=8.0cm]{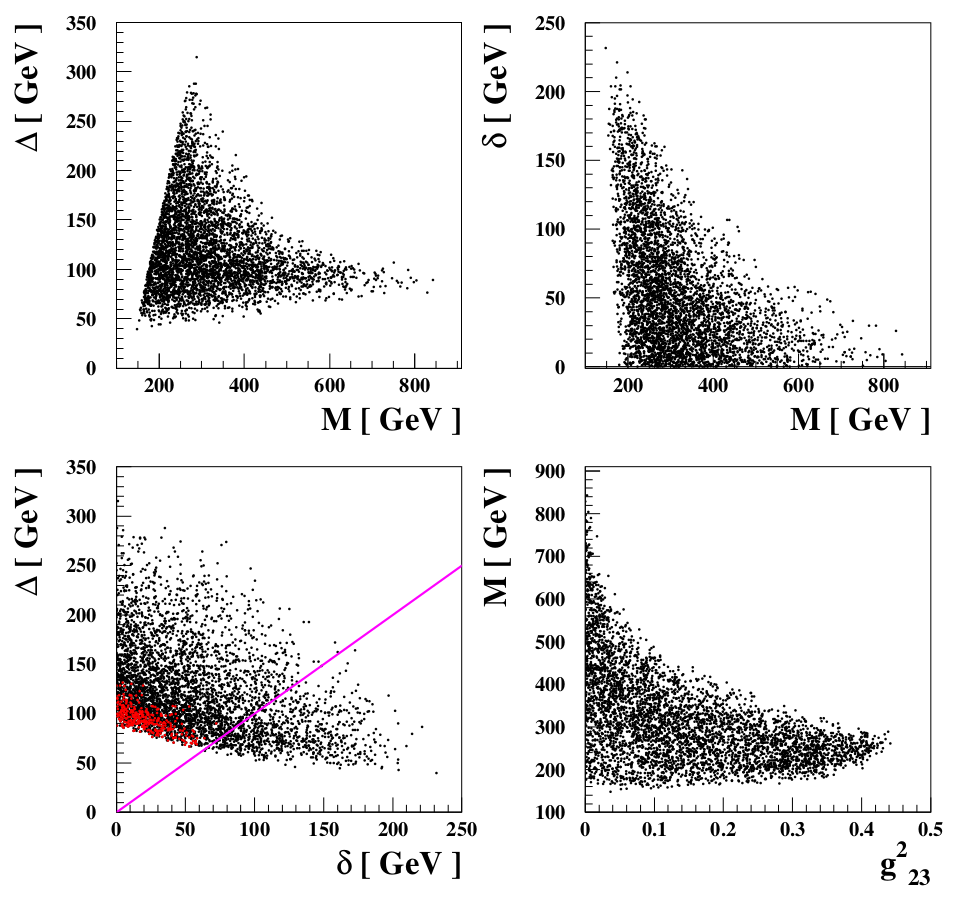}
\end{center}
\vspace{-0.5cm}
\caption{\it {\bf CDF}:
[Left] Scatter plots of
$\Delta$ versus $M$ (upper left),
$\delta$ versus $M$ (upper right),
$\Delta$ versus $\delta$ (lower left), and
$M$ versus $g_{23}^2$ (lower right) for
$\Delta <0$.
[Right] The same as in the left panel but for $\Delta >0$.
The magenta lines in the lower-left frames in each panel are for
$\Delta=-\delta$ (Left) and  $\Delta=\delta$ (Right).
And the red and blue dots are for
$M > 500$ GeV (red$+$blue) and $M > 900$ GeV (blue), respectively.
The combined UNIT$\oplus$BFB$\oplus$ELW$_{95\%}$ constraints are imposed.
}
\label{fig:ddd}
\end{figure}
In Fig.~\ref{fig:ddd}, we show the correlations among
$\Delta$, $\delta$, $M$, and $g_{23}^2$ imposing the combined
UNIT$\oplus$BFB$\oplus$ELW$_{95\%}$ constraints for
$\Delta<0$ (Left) and $\Delta>0$ (Right).
First of all, we observe that $M$ is constrained from the above
and $|\Delta|$ is  non-vanishing and sizable:
$M\lsim 1050$ GeV and
$M\lsim 850$ GeV when $\Delta<0$ and $\Delta>0$, respectively,
and $\Delta$ converges to the values
of about $\pm 90$ GeV as $M$ reaches its maximum values.
On the other hand, 
the mass splitting $\delta$ between $H_3$ and $H_2$ becomes smaller
as $M$ increases as shown in Eq.~(\ref{eq:Z5delta})
and $H_3$ could be (almost) degenerate with $H_2$
for the whole range of $M$.
We further observe that $|\Delta|>\delta$ when $M>550$ GeV, see the 
colored dots below (Left) and above (Right) the magenta lines
for $|\Delta|=\delta$ in the lower-left frames.
The scatter plots for $M$ versus $g_{23}^2$ in the lower-right frames
illustrate the effects of deviation from the alignment limit.
The upper limit on $M$ quickly decreases as $g_{23}^2$
deviates from the alignment limit and we find
find $M\lsim 700\,(600)$ GeV and $M\lsim 600\,(500)$ GeV 
when $g_{23}^2=0.05\,(0.1)$.
In this work, we concentrate on
the heavy mass region with $M\gsim 500$ GeV where $|\Delta|\gsim\delta$ 
and, accordingly, there is a mass  hierarchy such as
\begin{equation}
\label{eq:hierarchy}
M\gg M_{H_1}\sim |\Delta| \gsim \delta\,,
\end{equation}
which leads to the following two types of mass spectrum
\begin{eqnarray}
\label{eq:h2lchl}
{\bf 23C}: \ \ \
&&
M_{H_2}\ <\ M_{H_3}\ < \ M_{H^\pm} \ \ \ {\rm when} \ \ \ \Delta<0 \ \ \
{\rm and} \ \ \ \delta < |\Delta|\,; \nonumber \\[2mm]
{\bf C23}: \ \ \
&&
M_{H^\pm}\ <\ M_{H_2}\ < \ M_{H_3} \ \ \ {\rm when} \ \ \ \Delta>0 \,. 
\end{eqnarray}
For the 23C hierarchy, the second heaviest Higgs boson $H_2$ is the lightest state 
while the charged Higgs boson is the lightest one for the C23 hierarchy.
The heaviest states are $H^\pm$ and $H_3$ for the 23C and C23 hierarchies, respectively.

Assuming the mass hierarchy given in Eq.~(\ref{eq:hierarchy}) and keeping the 
leading terms for $T$ and $S$ using the approximations 
given in Eqs.~(\ref{eq:fapp1}) and (\ref{eq:fapp2}),
one may find the approximated expressions for the $S$ and $T$ parameters:
\begin{eqnarray}
\label{eq:STapp}
\frac{T}{\frac{\sqrt{2}G_F}{16\pi^2\alpha_{\rm EM}}} &\simeq&
 \frac{4}{3}(\Delta^2+\delta\Delta)+\left[
-\left(2g_{23}^2\Delta+g_{_{H_2VV}}^2\delta\right)\,M
-\frac{4}{3}g_{23}^2\Delta^2
-\left(\frac{1}{2}g_{_{H_2VV}}^2+\frac{4}{3}g_{_{H_3VV}}^2\right)\delta\Delta
+\frac{1}{6}g_{_{H_2VV}}^2\delta^2\right] \,, \nonumber \\[2mm]
4\pi\,S &\simeq&
 \frac{1}{3}\left(2\frac{\Delta}{M}+\frac{\delta}{M}\right)
+\left[-\frac{5}{9}g_{23}^2
+\frac{1}{3}(g_{_{H_2VV}}^2-g_{_{H_3VV}}^2)\frac{\delta}{M}\right]\,,  
\end{eqnarray}
in terms of $\Delta$, $\delta$, $M$, $g_{_{H_2VV}}^2$, $g_{_{H_3VV}}^2$, and $g_{23}^2$.
Note that the quantities in the square brackets are vanishing in the alignment limit
of $g_{_{H_2VV}}^2=g_{_{H_3VV}}^2=g_{23}^2=0$ and the terms 
are organized taking account of the mass hierarchy 
$M\gg M_{H_1}\sim |\Delta| \gsim \delta$,  see Eq.~(\ref{eq:hierarchy}).
Further we observe that the terms containing
either $g_{_{H_2VV}}^2$ or $g_{_{H_3VV}}^2$
vanish when $\delta=0$ leaving only the terms
with the average coupling $g_{23}^2$.
One should note that $T$ is vanishing when $\Delta=\delta=0$ and one need mass 
splittings among the heavy Higgs bosons 
to accommodate the sizable positive value of the $T$ parameter.
When $\delta=0$, $T$ still have positive contributions proportional to 
$\Delta^2$ even in the alignment limit of $g_{23}^2=0$.
When $\Delta=0$, on the other hand, $T$ is vanishing in the alignment and
it is given by
\begin{equation}
\left.
\frac{T}{\frac{\sqrt{2}G_F}{16\pi^2\alpha_{\rm EM}}}\right|_{\Delta=0} \simeq
-g_{_{H_2VV}}^2 \delta \left( M -\frac{1}{6}\delta\right)\,,
\end{equation}
which is negative unless $\delta>6M$ or 
identically vanishing when $g_{_{H_2VV}}=0$ or when $H_2$ is the CP-odd state.
The large mass splitting between $H_3$ and $H_2$ such as $\delta>6M$
is not allowed since it generates too large value of $|Z_5|$
violating the perturbative UNIT constraint, see Eq.~(\ref{eq:Z1456align}),
let alone it is not consistent with our assumed mass hierarchy. 
\footnote{We also note that $\delta \sim 6M$ leads to
$S\sim 1/2\pi$ when $\Delta=g_{23}^2=0$, see Eq.~(\ref{eq:STapp}),
which is far outside the scattered regions presented in Fig.~\ref{fig:TversusS}
which are obtained by imposing
the combined UNIT$\oplus$BFB$\oplus$ELW$_{95\%}$ constraints.}
And, $T=0$ with $g_{_{H_2VV}}=0$ is consistent with the observation made 
in the CPC case~\cite{Heo:2022dey}: $T$ is identically vanishing when the mass of
the charged Higgs boson is equal to that of the CP-odd Higgs boson.
So we conclude that, only with non-zero $\delta$, we 
are heading to the opposite direction pointed by
the CDF measurement of $M_W$ and
non-vanishing $\Delta$ is necessary to produce the
sizable and positive definite value of the $T$ parameter.
To strengthen our claim, we extend our discussion to the case 
with $\Delta\neq 0$. In this case, 
dropping the terms suppressed by the factor $\Delta/M$ or $\delta/M$
compared to the leading term in the square bracket in Eq.~(\ref{eq:STapp}),
the expression for $T$ is further simplified into
\begin{equation}
\frac{T}{\frac{\sqrt{2}G_F}{16\pi^2\alpha_{\rm EM}}} \simeq
 \frac{4}{3}(\Delta^2+\delta\Delta)
-\left(2g_{23}^2\Delta+g_{_{H_2VV}}^2\delta\right)\,M\,, 
\end{equation}
which can be rewritten as
\begin{equation}
\left[\Delta-\left(\frac{3}{4}g_{23}^2\,M-\frac{\delta}{2}\right)\right]^2 \simeq
T^\prime + \frac{3}{4}g_{_{H_2VV}}^2\delta\,M +
\left(\frac{3}{4}g_{23}^2\,M-\frac{\delta}{2}\right)^2  
\end{equation}
with $T^\prime \equiv \frac{T}{0.18}\,\left(100\,{\rm GeV}\right)^2$
using $\sqrt{2}G_F/16\pi^2\alpha_{\rm EM} \simeq 1.337\times 10^{-5}/{\rm GeV}^2$.
Note that the right-hand side of the above equation is positive definite unless $T<0$,
$\Delta=0$ cannot satisfy the above relation,
and the region of $\Delta$ with a radius smaller than $\sqrt{T^\prime}$ around
the point $\Delta=3g_{23}^2\,M/4-\delta/2$  should not be allowed.
For example, if $T\gsim 0.18$,
the region of $|\Delta| \lsim  100$ GeV should be excluded
in the limit of $\delta=g_{23}^2=0$.
In fact, we have checked that $T$ is negative definite when $\Delta=0$
using the full analytic expressions for the $T$ parameter
without resorting to the approximations, 
see Eqs.~(\ref{eq:fapp1}) and (\ref{eq:fapp2}). 
\footnote{
In fact, $T$ is identically vanishing when $\Delta=\delta=0$. 
We find that the same thing happens also when $\Delta=M_{H_2}-M_{H_1}=0$.}

As shown in our previous work~\cite{Heo:2022dey},
the quartic coupling $Z_4$ plays the crucial role to understand
the upper limit on the mass scale of heavy Higgs bosons. 
We find that the expression for $Z_4$ in Eq.~(\ref{eq:Z1456}) can be recast as
\begin{equation}
\label{eq:Z4}
Z_4 v^2 =
 2(2\Delta+\delta) M +(\Delta+\delta)\delta -\left[
 2g_{23}^2 M^2
+2(g_{23}^2 \Delta  +g_{_{H_3VV}}^2 \delta ) M
+g_{23}^2 \left(\frac{\Delta^2}{2}-2 M_{H_1}^2\right)
+g_{_{H_3VV}}^2(\Delta+\delta)\delta\right]\,,
\end{equation}
in terms of $\Delta$, $\delta$, $M$, 
$g_{23}^2$.  $g_{_{H_3VV}}^2$, and $M_{H_1}$.
When $g_{23}^2=0$ and $\delta=0$, using the above equation together with Eq.~(\ref{eq:STapp}), 
we have 
\begin{eqnarray}
T &\simeq & 0.18\,\left(\frac{\Delta}{100\,{\rm GeV}}\right)^2 \ \ \ {\rm or} \ \ \
|\Delta | \simeq 100\,\sqrt{\frac{T}{0.18}}\,{\rm GeV}\,, \nonumber \\[2mm]
S &\simeq & 0.0053\,
\left(\frac{\Delta}{100\,{\rm GeV}}\right)\,
\left(\frac{1\,{\rm TeV}}{M}\right)\,, \nonumber \\[2mm]
Z_4 &\simeq & 4 \frac{\Delta\,M}{v^2} \simeq 6.6\,
\left(\frac{\Delta}{100\,{\rm GeV}}\right)\,
\left(\frac{M}{1\,{\rm TeV}}\right)\,,
\end{eqnarray}
where we use
$\sqrt{2}G_F/16\pi^2\alpha_{\rm EM} \simeq 1.337\times 10^{-5}/{\rm GeV}^2$
and $v=246.22$ GeV.
Note that the UNIT$\oplus$BFB constraint of
$|Z_4|\lsim 6.6$  leads to the upper limit of about $1$ TeV on 
the heavy-Higgs mass scale $M$
when $|\Delta|\simeq 100$ GeV. 
These are the main ingredients of our previous work~\cite{Heo:2022dey}.

In this work, we extend our analytic study on the mass spectrum of heavy Higgs bosons
taking account of the effects of the mass splitting $\delta$ between the two 
neutral heavy Higgs bosons and the deviation from the
alignment limit of $g_{23}^2=0$.
Assuming the mass hierarchy $M\gg M_{H_1}\sim |\Delta| \gsim \delta$
given by Eq.~(\ref{eq:hierarchy}) and keeping the leading terms in Eq.~(\ref{eq:Z4}), 
one might approximate the expression for $Z_4$ as follow:
\begin{eqnarray}
\label{eq:Z4app}
Z_4 v^2 &\simeq &
 \left(4\Delta+2\delta\right)\, M -2g_{23}^2 M^2\,.
\end{eqnarray}
It is interesting and illuminating to
observe that each term of the above equation involves its own physics origin.
The term containing $Z_4$ is constrained by the UNIT$\oplus$BFB condition.
The first term in the right-hand side involves the mass splittings among
the heavy Higgs bosons and the high-precision CDF measurement of 
the $W$ boson mass implies non-vanishing value for, especially, the
mass splitting $\Delta$ between the charged and the 
second heaviest neutral Higgs bosons.
The second term in the right-hand side is proportional to $M^2$ and it 
quickly increases when $g_{23}^2$ deviates from the alignment limit of $g_{23}^2=0$
or when the coupling $g_{_{H_1VV}}$ deviates from its SM value of $1$.

Eq.~(\ref{eq:Z4app}) can be rewritten for $M$ as
\begin{equation}
\label{eq:MSQ}
g_{23}^2 M^2 -2\widehat\Delta\,M+2\widehat\Delta\,\widehat{M}_0\simeq 0\,,
\end{equation}
by introducing
\begin{equation}
\widehat\Delta \equiv \Delta +\frac{\delta}{2}\,, \ \ \
\widehat{M}_0 \equiv  \frac{Z_4v^2}{4\Delta+2\delta} = \frac{Z_4v^2}{4\widehat\Delta}\,.
\end{equation}
When the discriminant
$-2g_{23}^2\widehat\Delta\widehat{M}_0+\widehat\Delta^2>0$,
Eq.~(\ref{eq:MSQ}) can be solved for $M$. By denoting the solution as $M_{Z_4}$,
we obtain:
\footnote{Note that $\widehat\Delta$ has the same sign as $\Delta$ 
and $\widehat{M}_0>0$ when
$M\gsim 500$ GeV and $|\Delta|\gsim \delta$.}
\begin{eqnarray}
\label{eq:MZ4}
&& {\bf 23C}\,: \quad
M_{Z_4} \simeq \frac{|\widehat\Delta|}{g_{23}^2}\left[ -1 +
\left(1-2g_{23}^2\frac{\widehat{M}_0}{\widehat\Delta}\right)^{1/2}\right] \ \ \
{\rm when} \ \  \widehat\Delta<0 \ \ {\rm and} \ \
2g_{23}^2\widehat{M}_0>-|\widehat\Delta| \,,
\nonumber \\[2mm]
&& {\bf C23}\,: \quad
M_{Z_4} \simeq \frac{\widehat\Delta}{g_{23}^2}\left[ 1-
\left(1-2g_{23}^2\frac{\widehat{M}_0}{\widehat\Delta}\right)^{1/2}\right] \ \ \
{\rm when} \ \  \widehat\Delta>0 \ \ {\rm and} \ \
2g_{23}^2\widehat{M}_0<\widehat\Delta\,.
\end{eqnarray}

When $g_{23}^2|\widehat{M}_0/\widehat\Delta|\lsim 1$,
and one may expand $M_{Z_4}$ as
\begin{equation}
\label{eq:MZ4app}
M_{Z_4} \simeq \widehat{M}_0
\left[1+\frac{g_{23}^2}{2}\frac{\widehat{M}_0}{\widehat\Delta}
+\frac{g_{23}^4}{2}\left(\frac{\widehat{M}_0}{\widehat\Delta}\right)^2
+{\cal O}\left(g_{23}^2\frac{\widehat{M}_0}{\widehat\Delta}\right)^3 \right]
\end{equation}
and, keeping the leading term in $g_{23}^2$, we define
\begin{equation}
\label{eq:MZ4app1}
M_{Z_4}^{(1)} \equiv \widehat{M}_0
\left[1+\frac{g_{23}^2}{2}\frac{\widehat{M}_0}{\widehat\Delta}\right]\,.
\end{equation}
Note that the quartic coupling $Z_4$ and the 
quantities $\widehat\Delta=\Delta+\delta/2$ 
and $\widehat{M}_0=Z_4v^2/4\widehat\Delta$ possess their own
{\it implicit} dependence on $g_{23}^2$
as will be shown shortly.

\begin{figure}[t!]
\vspace{-0.5cm}
\begin{center}
\includegraphics[width=8.0cm]{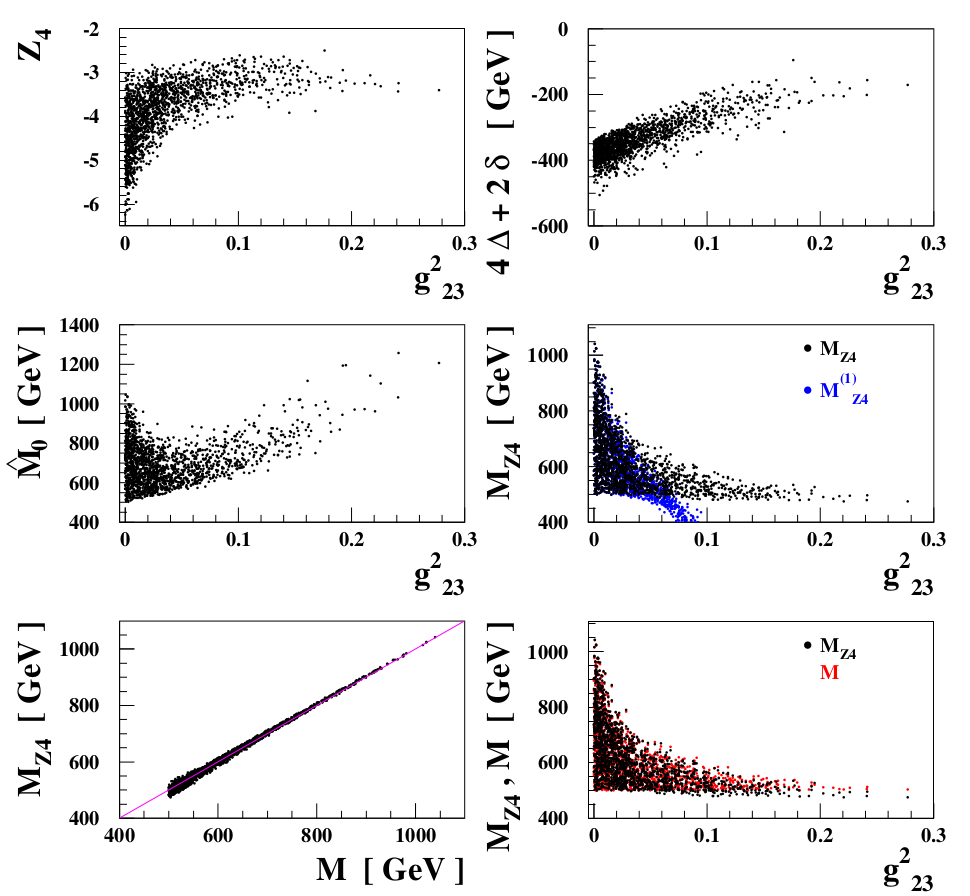}
\includegraphics[width=8.0cm]{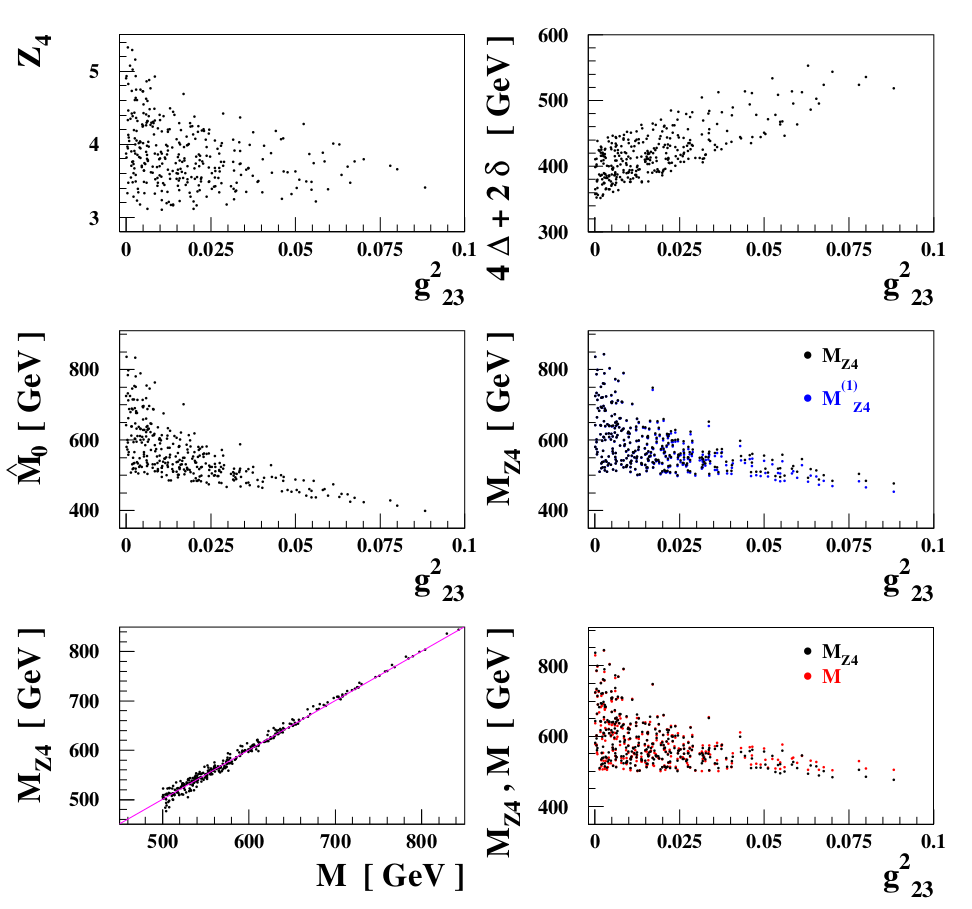}
\end{center}
\vspace{-0.5cm}
\caption{\it {\bf CDF}:
Behaviors of $Z_4$ (upper-left), $4\Delta+2\delta$ (upper-right),
$\widehat{M}_0$ (middle-left), and
$M_{Z_4}$ and $M_{Z_4}^{(1)}$ (middle-right) according to the variation of $g_{23}^2$
and comparisons of $M_{Z_4}$ and $M$ (lower)
for the {\bf 23C} (Left) and {\bf C23} (Right) hierarchies with the magenta lines for $M_{Z_4}=M$.
Note that $M_{Z_4}$ and $M_{Z_4}^{(1)}$ 
are given by Eq.~(\ref{eq:MZ4}) and Eq.~(\ref{eq:MZ4app1}),
respectively. 
The combined UNIT$\oplus$BFB$\oplus$ELW$_{95\%}$ constraints are imposed
and  $M>500$ GeV is required.
}
\label{fig:m0h}
\end{figure}
In the left panel of Fig.~\ref{fig:m0h}, we show the scatter plots for
$Z_4$ versus $g_{23}^2$ (upper left),
$4\Delta+2\delta$ versus $g_{23}^2$ (upper right),
$\widehat{M}_0$  versus $g_{23}^2$ (middle left), and
$M_{Z_4}$ and $M_{Z_4}^{(1)}$  versus $g_{23}^2$ (middle right)
for the 23C hierarchy with $\Delta<0$.
The combined UNIT$\oplus$BFB$\oplus$ELW$_{95\%}$ constraints 
have been imposed and $M>500$ GeV is required.
We observe that $|Z_4|$ takes its maximum value of about $6.3$
at $g_{23}^2=0$ where $350\lsim |4\Delta+2\delta|/{\rm GeV} \lsim 450$
which leads to $\widehat{M}_0=Z_4v^2/(4\Delta+2\delta)\lsim 1050$ GeV in the alignment limit.
As $g_{23}^2$ deviates from 0, the maximum value of $|Z_4|$ decreases quickly 
from $\sim 6.3$ to $\sim 4$ as $g_{23}^2$ reaches to $0.05$ and, from there,
it does not change much.
On the other hand, the minimum value of $|4\Delta+2\delta|$ 
almost uniformly decreases as $g_{23}^2$ increases.
These combined features
explain the behavior of $\widehat{M}_0$ shown the middle-left frame.
In the middle-right frame, we show $M_{Z_4}$ and $M_{Z_4}^{(1)}$ which 
are given by Eq.~(\ref{eq:MZ4}) and Eq.~(\ref{eq:MZ4app1}),
respectively. We observe that, for the small values of $g_{23}^2\lsim 0.05$, 
$M_{Z_4}$ is fairly well approximated by
$M_{Z_4}^{(1)}$ which contains the proportionality factor
$1+g_{23}^2\widehat{M}_0/2\widehat\Delta$ 
pulling down $\widehat{M}_0$ as $g_{23}^2$ increases since
$\widehat\Delta<0$ for the 23C hierarchy.
In the lower frames, we compare $M_{Z_4}$ obtained by solving Eq.(\ref{eq:MSQ})
and the average mass scale $M=(M_{H^\pm}+M_{H_2})/2$. 
We find that $M_{Z_4}\simeq M$ with $|M_{Z_4}-M|\lsim 30$ GeV and, therefore,
conclude that $M_{Z_4}$ excellently represents the heavy-Higgs mass scale $M$
especially for the small values of $g_{23}^2$.
This implies that we have fully figured out the physics origins 
relevant for the mass scale of heavy Higgs bosons by showing that it
has been determined by combining the UNIT$\oplus$BFB condition,
the mass splittings among the heavy Higgs bosons, and
deviation from the alignment limit.

In the right panel of Fig.~\ref{fig:m0h}, we show the same as in the left
panel  but for the C23 hierarchy with $\Delta>0$.
Similar observations could be made as in the left panel 
except that: $(i)$ $g_{23}^2$ is constrained to be smaller than about $0.1$
indicated by the condition $2g_{23}^2 \widehat{M}_0<\widehat\Delta$,
$(ii)$ $Z_4\lsim 5.3$ 
leading to the smaller values of $\widehat{M}_0\lsim 850$ GeV, 
$(iii)$ now $\widehat{M}_0$ is pushed up by the factor
$1+g_{23}^2\widehat{M}_0/2\widehat\Delta$ since $\widehat\Delta>0$, and
$(iv)$ $M_{Z_4}^{(1)}$ provides the excellent approximation for $M_{Z_4}$.
Otherwise, we confirm again that
we have fully figured out the physics origins 
relevant for the mass scale of heavy Higgs bosons 
also in this case.

To conclude, for a given value of $g_{23}^2$,
the upper limit on the average
mass scale $M=(M_{H^\pm}+M_{H_2})/2$ is given by the maximum value of $M_{Z_4}$
which could be approximated by $M_{Z_4}^{(1)}$ when $g_{23}^2\lsim 0.05$:
\begin{equation}
M
\lsim {\rm max}\left(M_{Z_4}\right) 
\simeq {\rm max}\left[M_{Z_4}^{(1)}\right]\,,
\end{equation}
and then
\begin{equation}
M_{H_2} = M +\Delta/2 \lsim {\rm max}\left[M_{Z_4}^{(1)}\right] +\Delta/2\,,  \ \ \
M_{H^\pm} = M -\Delta/2 \lsim {\rm max}\left[M_{Z_4}^{(1)}\right] -\Delta/2\,,
\end{equation}
for the both 23C ($\Delta<0$) and C23 ($\Delta>0$) hierarchies.
In this section,
taking the CDF values 
given in Eq.~(\ref{eq:STEXP}), we find
\begin{equation}
{\rm max}\left[M_{Z_4}^{(1)}\right] =
{\rm max}\left(\widehat{M}_0\right) \simeq \left\{
\begin{array}{rcc}
1050\ {\rm GeV} & {\rm for} & {\bf 23C}\,(\Delta<0)\\[2mm]
850\ {\rm GeV}  & {\rm for} & {\bf C23}\,(\Delta>0)
\end{array}\right.\,,
\end{equation}
in the alignment limit which precisely reproduces the highest masses of the heavy 
neutral and charged Higgs bosons shown in Fig.~\ref{fig:mmcorr} 
together with $\left|\Delta\right|_{M={\rm max}(M)} \simeq 100$ GeV, 
see Eq.~(\ref{eq:mnc}).
Note that the 23C hierarchy leads to the heavier spectrum.

\section{Discussion}
\label{sec:discussion}
The specific values for the $S$ and $T$ parameters given in Eq.~(\ref{eq:STEXP}) 
are subject to changes depending on the details of 
the available ELW precision data and how to collect and fit them.
In this section, taking PDG and CDF as the two extreme limits, we discuss
how the upper limit on the heavy-Higgs mass scale $M$ changes by varying
the values for the $S$ and $T$ parameters between PDG and CDF.
More precisely, we vary the central values of the $S$ and $T$ parameters as
\begin{equation}
\label{eq:s0t0}
\widehat{T}_0=0.22\,t_{EW} + 0.05\,, \ \ \
\widehat{S}_0=0.15\,t_{EW}\,
\end{equation}
with the parameter $t_{EW}$ scanned between 0 and 1.
For $\sigma_S$, $\sigma_T$, and $\rho_{ST}$, we take the CDF values as
in Eq.~(\ref{eq:STEXP}).
Now, the heavy Higgs masses squared are scanned up to $(3~{\rm TeV})^2$
since we expect the smaller mass splitting $|\Delta|$ with
$\widehat{T}_0<0.27$ which might lead to the upper limit on $M$ larger than $1$ TeV. 
Further we take $\gamma=\eta=0$ because
$M_{Z_4}$ takes its maximum value in the alignment limit, see  
the middle frames in Fig.~\ref{fig:m0h}.
Otherwise, the parameters are
scanned as described at the end of Section \ref{sec:framework}.

\begin{figure}[t!]
\vspace{-0.5cm}
\begin{center} 
\includegraphics[width=8.0cm]{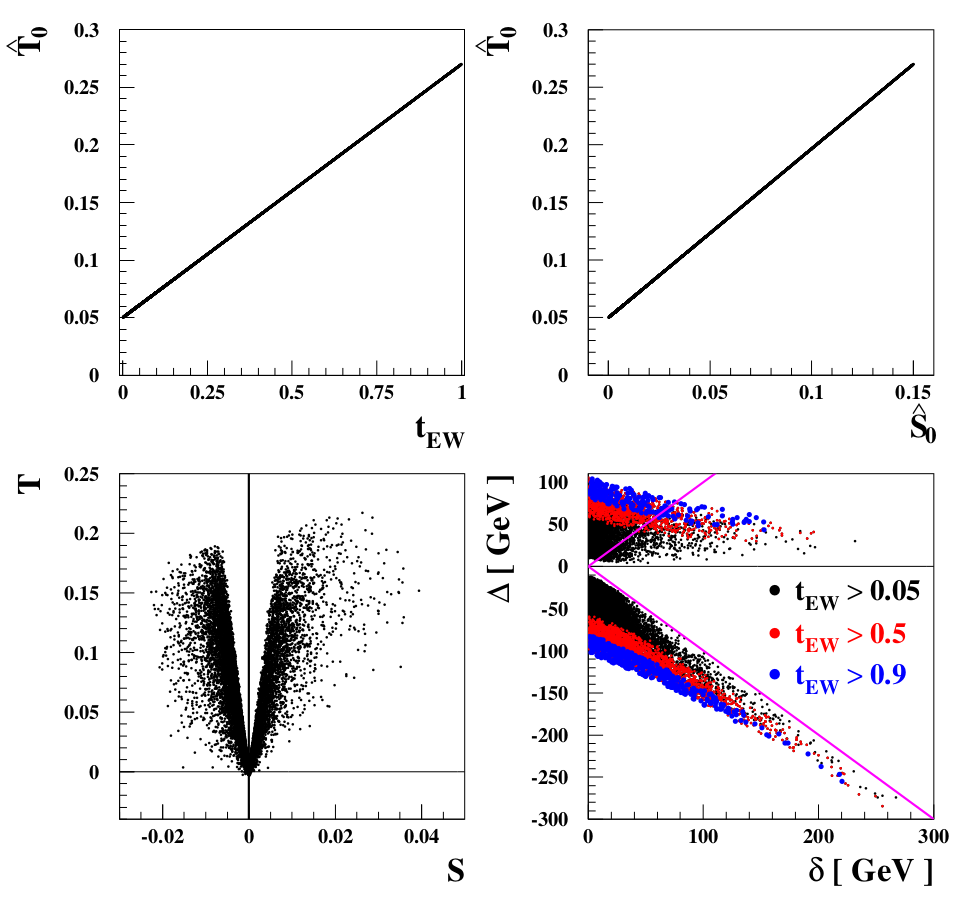}
\includegraphics[width=8.0cm]{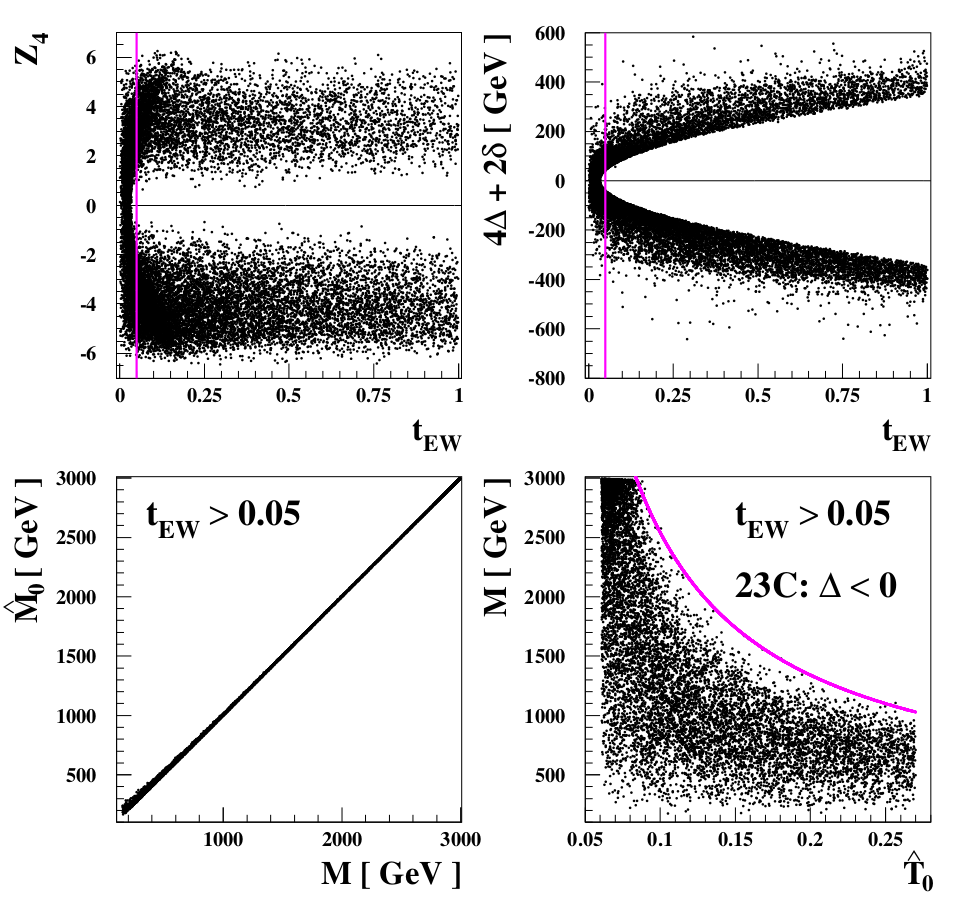}
\end{center}
\vspace{-0.5cm}
\caption{\it Between {\bf PDG} and {\bf CDF}:
$\widehat{T}_0=0.22\,t_{EW} + 0.05$ and
$\widehat{S}_0=0.15\,t_{EW}$ with $0\leq t_{EW} \leq 1$ taking 
the CDF values for $\sigma_S=0.08$, $\sigma_T=0.06$, and $\rho_{ST}=0.93$.
The alignment limit of $\gamma=\eta=0$ has been taken,
the heavy Higgs masses squared are scanned up to $(3~{\rm TeV})^2$, and
the other parameters are
scanned as described at the end of Section \ref{sec:framework}.
The combined UNIT$\oplus$BFB$\oplus$ELW$_{95\%}$
constraints have been imposed. 
(Left)
The magenta lines in the lower-right frame 
are for $|\Delta|=\delta$ and the colored dots are for
$t_{EW}>0.05$ (black$+$red$+$blue),
$t_{EW}>0.5$ (red$+$blue), and
$t_{EW}>0.9$ (blue).
(Right)
The vertical magenta lines in the upper frames 
denotes the position $t_{EW}=0.05$ and
the magenta curve in the lower-right frame 
represents the case $M=240/\widehat{T}_0+140$ GeV.
}
\label{fig:s0t0}
\end{figure}

The left panel of Fig.~\ref{fig:s0t0} are for the scatter plots for
$\widehat{T}_0$ versus $t_{EW}$ (upper left) and
$\widehat{T}_0$ versus $\widehat{S}_0$ (upper right) according to
Eq.~(\ref{eq:s0t0}). 
The lower-left frame is for $T$ versus $S$ 
obtained by imposing the combined UNIT$\oplus$BFB$\oplus$ELW$_{95\%}$
constraints.
Comparing with Fig.~\ref{fig:TversusS}, we observe that
the scattered dots are more centered around $S=0$ with $T>0$.
The lower-right frame is for $\Delta$ versus $\delta$ 
with the magenta lines for $|\Delta|=\delta$. 
The colored dots are for
$t_{EW}>0.05$ (black$+$red$+$blue),
$t_{EW}>0.5$ (red$+$blue), and $t_{EW}>0.9$ (blue).
First of all,
we note that the non-vanishing mass splitting $|\Delta|$ 
is required if $\widehat{T}_0$ deviates from 0 by more
than $1\sigma_T=0.06$ or, equivalently, $t_{EW}>0.05$.
And, as expected, the larger mass splitting $|\Delta|$ is needed
to accommodate the larger value of $\widehat{T}_0$. 
We note that $|\Delta|\simeq 100$ GeV when $\delta=0$
in the CDF limit of $t_{EW}=1$.
When $\Delta<0$, $|\Delta|>\delta$ resulting in the 23C hierarchy 
independently of $t_{EW}$ and $M$.

In the upper frames of the right panel of Fig.~\ref{fig:s0t0}, 
we show the scatter plots for
$Z_4$ versus $t_{EW}$ (upper left) and
$4\Delta+2\delta$ versus $t_{EW}$ (upper right).
From $t_{EW}\simeq 0.05$ denoted by the vertical
magenta lines, the region around $|Z_4|=0$ starts 
to be excluded and $|4\Delta+2\delta|$ begins to deviate from 0
implying that the upper limit on $M$ emerges.
Note that the allowed regions of the quartic coupling $Z_4$ 
are almost independent of $t_{EW}$ once it is larger than $0.05$.
In the lower-left frame, we show the scatter plot for
$\widehat{M}_0$ versus $M$ for $t_{EW}>0.05$. 
We observe that the scattered points are distributed along the
$\widehat{M}_0 = M$ line with 
$|\widehat{M}_0-M|<20$ GeV for $M>1$ TeV
\footnote{In the alignment of $g_{23}^2=0$,
$M_{Z_4}=\widehat{M}_0=Z_4v^2/(4\Delta+2\delta)$, see
Eq.~(\ref{eq:MSQ}).}
implying that the mass scale of heavy Higgs bosons originates from
the UNIT$\oplus$BFB constraint on the quartic coupling $Z_4$ and
the mass splittings among the heavy Higgs bosons.
In the lower-right frame, taking $t_{EW}>0.05$,
we show the scatter plot for $M$ versus $\widehat{T}_0$ for 
the 23C hierarchy ($\Delta<0$) which predicts
the heavier mass spectrum compared to the C23 hierarchy with $\Delta>0$.
We observe that the heavy Higgs bosons should exist
below $2.5\,(1.3)$ TeV if $\widehat{T}_0>0.1\,(0.2)$.
Incidentally, we find that
\begin{equation}
\label{eq:T0Mupperlimit}
M\ \lsim \ \left(\frac{240}{\widehat{T}_0}+140\right)\ {\rm  GeV}\,,
\end{equation}
as denoted by the magenta curve in the lower-right frame.
This empirical relation could be used to derive the 
conservative $\widehat{T}_0$-dependent upper limit on $M$ 
more conveniently. 

\newcolumntype{C}[1]{>{\centering\arraybackslash}p{#1}}
\setlength{\extrarowheight}{0.15cm}
%
\begin{table}[t]
\centering
\caption{\it 
\label{tab:benchmarking}
Benchmarking scenarios in which the masses of the heavy
Higgs bosons saturate their upper limits for a given value of $\widehat{T}_0$.
We take the four values of $\widehat{T}_0$
from the CDF value of $0.27$ to $0.09$
in decrement of $\sigma_T=0.06$
with $\widehat{S}_0=0.15\,(\widehat{T}_0-0.05)/0.22$,
see Eq.~(\ref{eq:s0t0}).}
\begin{tabular}{|C{2cm}|C{2cm} C{2cm}|C{2cm} C{2cm}|}
\hline
\multicolumn{1}{|c|}{\multirow{2}{*}{$(\widehat{T}_0\,,\widehat{S}_0)$}} & 
\multicolumn{2}{c|}{\bf 23C} & \multicolumn{2}{c|}{\bf C23}                         \\ \cline{2-5}
\multicolumn{1}{|c|}{} & \multicolumn{1}{c|}{$M_{H_2} \simeq M_{H_3}$ (GeV)} &
\multicolumn{1}{c|}{$M_{H^\pm}$ (GeV)} & \multicolumn{1}{c|}{$M_{H_2}\simeq M_{H_3}$ (GeV)} &
\multicolumn{1}{c|}{$M_{H^\pm}$ (GeV)} \\ \hline
\multicolumn{1}{|c|}
{$(0.27\,,0.15)$} & \multicolumn{1}{c|}{$1000$} & \multicolumn{1}{c|}{$1100$} &
\multicolumn{1}{c|}{$900$} & \multicolumn{1}{c|}{$800$} \\
{$(0.21\,,0.11)$} & \multicolumn{1}{c|}{$1100$} & {$1180$} & \multicolumn{1}{c|}{$1090$} & {$1010$} \\
{$(0.15\,,0.07)$} & \multicolumn{1}{c|}{$1640$} & {$1700$} & \multicolumn{1}{c|}{$1540$} & {$1480$} \\
{$(0.09\,,0.03)$} & \multicolumn{1}{c|}{$2730$} & {$2760$} & \multicolumn{1}{c|}{$2630$}& {$2600$} \\
\hline
\end{tabular}
\end{table}

Last but not least, 
for the searches of heavy Higgs bosons at future colliders such as
the high luminosity option of the LHC (HL-LHC) and
a 100 TeV hadron collider,
we propose some benchmarking scenarios 
in which the masses of the heavy Higgs bosons saturate their 
upper limits for a given value of $\widehat{T}_0$,
see Table~\ref{tab:benchmarking}.
In the proposed saturating scenarios, the heavy neutral Higgs bosons
are almost degenerate.
The heavier spectrum is predicted for
the larger value of $\widehat{T}_0$ and we note that
the heavy Higgs bosons are lighter than 3 TeV
for $\widehat{T}_0>0.09$.
When $\widehat{T}_0\gsim 0.21$,
the heavy Higgs bosons weighing around 1 TeV 
and the on-shell $W$ boson coming from the decay of 
the charged Higgs boson (23C) or
the heavy neutral Higgs bosons (C23) might be signatures of
the scenarios with $|g_{_{H_{2,3}H^\pm W^\mp}}|\simeq 1$~\cite{Choi:2021nql}.
For $\widehat{T}_0\lsim 0.15$, the
heavy Higgs bosons are located above $1.5$ TeV and the mass difference
between the charged and neutral Higgs bosons is smaller than $60$ GeV.
They might be out of the reach of the HL-LHC but well within the reach of
a 100 TeV hadron collider~\cite{Hajer:2015gka,Arkani-Hamed:2015vfh}.

\section{Conclusions}
\label{sec:conclusions}
We analyze the implication of the large positive deviation of the
$S$ and $T$ parameters from the SM values of zero
indicated  by the high-precision CDF
measurement of the $W$ boson mass
on the mass spectrum of heavy Higgs bosons
considering the most general CP-violating 2HDM potential.
We show that the mass splitting between the charged Higgs boson 
and the second heaviest neutral one is necessary
to accommodate the sizable positive value of the $T$ parameter especially.
Combining with the theoretical constraints on the quartic couplings
from the perturbative unitarity and for the Higgs potential to 
be bounded from below, we provide the comprehensive analysis 
why the masses of the heavy Higgs bosons should be bounded 
from above.

\smallskip
We further suggest the following points as the main results of this work:
\begin{enumerate}
\item 
We clearly demonstrate that there exist only two rephasing-invariant
CPV phases of $\theta_1={\rm Arg}[Z_6 (Z_5^*)^{1/2}]$ and
$\theta_2={\rm Arg}[Z_7 (Z_5^*)^{1/2}]$ in the CPV 2HDM Higgs potential
pivoting around the complex quartic coupling $Z_5$.
Note that the second angle $\theta_2$ has nothing to do with
the masses of Higgs bosons and the mixing of the neutral ones.
\item 
By imposing the combined UNIT$\oplus$BFB$\oplus$ELW$_{95\%}$
constraints  and taking the CDF values for the $S$ and $T$ parameters,
we find
\begin{eqnarray}
&&
0.1\lsim Z_1 \lsim 2.0\,, \ \ \
0  \lsim Z_2 \lsim 2.1\,, \ \ \
-2.6 \lsim Z_3 \lsim 8.0\,, \nonumber \\[2mm]
&& -6.3 \lsim Z_4 \lsim -0.8 \ \cup \
1.1 \lsim Z_4 \lsim 5.4\,,
\nonumber \\[2mm]
&&
|Z_5|\lsim 1.7\,, \ \ \
|Z_6|\lsim 2.4\,, \ \ \
|Z_7|\lsim 2.7\,. \nonumber
\end{eqnarray}
Note that the ranges of the quartic couplings appearing in the Higgs potential
are largely independent of the ELW constraint except that
the region with $|Z_4|\lsim 1$ is
excluded by the large value of the $T$ parameter.
\item
Concentrating on the heavy mass region $M=(M_{H_2}+M_{H^\pm})/2\gsim 500$ GeV in which 
we find a mass  hierarchy of 
$M\gg M_{H_1}\sim |\Delta| \gsim \delta$
with $\Delta =M_{H_2}-M_{H^\pm}$ and
$\delta = M_{H_3}-M_{H_2}$,
we figure out that there could be the following two types of mass spectrum
\begin{eqnarray}
{\bf 23C}: \ \ \
&&
M_{H_2}\ <\ M_{H_3}\ < \ M_{H^\pm} \ \ \ {\rm when}  \ \ \Delta<0 \ \ 
{\rm and}  \ \ \delta < |\Delta|\,; \nonumber \\[2mm]
{\bf C23}: \ \ \
&&
M_{H^\pm}\ <\ M_{H_2}\ < \ M_{H_3} \ \ \ {\rm when}  \ \ \Delta>0  \ \
{\rm independently\ of} \ \ \delta\,.\nonumber
\end{eqnarray}
When $\Delta<0$, note that one can achieve the hierarchy
$M_{H_3} <  M_{H^\pm}$ if $M\gsim 500$ GeV or
in the alignment limit.
\item
The upper limit on the masses of heavy 
Higgs bosons becomes stronger when 
the effects of deviation from the alignment limit are taken into account.
In the CDF case, we show that the upper limit on $M$ changes from about
$1$ TeV to about $700$ GeV as $g_{23}^2$ deviates from $0$ to $0.05$
for the 23C hierarchy.
\item
We have fully figured out the physics origins relevant for the mass scale 
of heavy Higgs bosons by showing that 
the heavy-Higgs mass scale $M$ could be excellently represented by
$M_{Z_4}$ obtained by solving Eq.~(\ref{eq:MSQ}) which incorporates.
$(i)$ the UNIT$\oplus$BFB condition,
$(ii)$ the mass splittings among the heavy Higgs bosons, and
$(iii)$ the deviation from the alignment limit.
\item
When $M\lsim 3$ TeV,
we provide the following useful and convenient
empirical $\widehat{T}_0$-dependent upper limit:
$$
M\ \lsim \ \left(\frac{240}{\widehat{T}_0}+140\right)\ {\rm  GeV}\,,
$$
obtained by varying the central values of the $S$ and $T$ parameters
between PDG and CDF given by Eq.~(\ref{eq:STEXP}).
\item 
We propose some $\widehat{T}_0$-dependent benchmarking scenarios in which
the masses of the heavy Higgs bosons saturate their upper limits
for a given value of $\widehat{T}_0$,
see Table~\ref{tab:benchmarking}.
For the CDF values given in Eq.~(\ref{eq:STEXP}), we propose
the following two {\tt sat-CDF} scenarios:
\begin{eqnarray}
{\bf 23C}_{\rm\tt sat-CDF}: \ \ \ &&
M_{H_2}\simeq M_{H_3} \simeq 1000\,{\rm GeV}\,, \ \ \ 
M_{H^\pm}\simeq 1100\,{\rm GeV}\,; \ \ \
\nonumber \\
{\bf C23}_{\rm\tt sat-CDF}: \ \ \ && M_{H^\pm}\simeq 800\,{\rm GeV}\,, \ \ \
M_{H_2} \simeq M_{H_3}\simeq 900\,{\rm GeV}\,.
\nonumber
\end{eqnarray}
\end{enumerate}

%
%
\section*{Acknowledgment}
We thank Chan Beom Park for the helpful comments 
regarding the numerical analysis.
This work was supported by the National Research Foundation (NRF) of Korea
Grant No. NRF-2021R1A2B5B02087078 (D.-W.J., Y.H., J.S.L.).
The work of D.-W.J. was also supported in part by
the NRF of Korea Grant Nos. NRF-2019R1A2C1089334 and 
NRF-2021R1A2C2011003 and in part by the Yonsei University Research Fund
of 2022.
The work of J.S.L. was also supported in part by
the NRF of Korea Grant No. NRF-2022R1A5A1030700.


\end{document}